\documentclass[a4paper,11pt]{article}
\pdfoutput=1
\usepackage{graphicx}
\usepackage{bm}
\usepackage{epsf}
\usepackage{rotating}
\usepackage{epsfig,graphics,rotate,color}
\usepackage{color}
\usepackage{array}
\usepackage[utf8]{inputenc}
\usepackage[T1]{fontenc}
\usepackage[caption=false]{subfig}
\usepackage[bottom]{footmisc}

\pdfoutput=1 % if your are submitting a pdflatex (i.e. if you have
\usepackage{jinstpub} % for details on the use of the package, please
\newcommand{\beq}{\begin{eqnarray}}
\newcommand{\eeq}{\end{eqnarray}}

\title{\boldmath Pulse Shape Particle Identification by a Single Large Hemispherical Photo-Multiplier Tube}

\author[a]{S.~Samani,\footnote{Now at University of Oxford, Oxford, OX1 3PU, UK.}}
\author[a]{S.~Mandalia,}
\author[b]{C.~Arg\"uelles,}
\author[b]{S.~Axani,}
\author[c]{Y.~Li,}
\author[b]{M.H.~Moulai,}
\author[d,e]{B.~Ty,}
\author[c]{and Z.~Xie,}
\author[b]{J.~Conrad,}
\author[a]{T.~Katori,\footnote{Corresponding author. Now at  King's College London, London, WC2R 2LS, UK.}}
\author[e]{P.~Sandstrom}

\affiliation[a]{School of Physics and Astronomy, Queen Mary University of London, London E1 4NS, UK}
\affiliation[b]{Department of Physics, Massachusetts Institute of Technology, Cambridge, MA 02139, USA}
\affiliation[c]{School of Physics, Sun Yat-sen University, Guangzhou, Guangdong, 510275 P. R. China}
\affiliation[d]{Department of Physics, University of Wisconsin, Madison, WI 53706, USA}
\affiliation[e]{Wisconsin IceCube Particle Astrophysics Center, Madison, WI 53706, USA}

\emailAdd{katori@fnal.gov}

\abstract{In neutrino experiments, hemispherical photomultiplier tubes (PMTs) are often used to cover large surfaces or volumes to maximize the photocathode coverage with a minimum number of channels. Instrumentation is often coarse, and neutrino event reconstruction and particle identification (PID) is usually done through the morphology of PMT hits. In future neutrino experiments, it may be desirable to perform PID from a few hits, or even a single hit, by utilizing pulse shape information. In this report, we study the principle of pulse shape PID using a single 25.4~cm hemispherical PMT in a spherical glass housing for future neutrino telescopes. We use the Fermilab Test Beam Facility (FTBF) MTest beam line to demonstrate that with pulse shape PID, it is possible to statistically separate 2~GeV electrons from 8~GeV pions, where the total charge deposition is \textasciitilde20~PE in our setup. Such techniques can be applied to future neutrino telescopes focusing on low-energy physics, including the IceCube-Upgrade.}


\begin{document}
\maketitle
\flushbottom

\newpage
\section{Introduction\label{sec:intro}}

  \subsection{The IceCube experiment\label{subsec:IC}}
    The IceCube Neutrino Observatory is a neutrino telescope located at the geographic South Pole, Antarctica~\cite{Aartsen:2016nxy}. It consists of an array of 5,160 Digital Optical Modules (DOMs) distributed and embedded in the natural glacial ice, spanning a total volume of \textasciitilde1km$^3$. Each DOM contains a downward-facing photomultiplier tube (PMT) in a spherical glass housing which detects Cherenkov photons emitted from high-energy charged particles traversing the ice, such as those from neutrino interactions, making the entire IceCube array volume a giant neutrino telescope.

    In the inner most part of the IceCube array, there is a region where DOMs are distributed with a higher density. These centrally located DOMs make up the subarray DeepCore~\cite{Collaboration:2011ym} region, which is sensitive to lower energy neutrinos down to a few~GeV. In the future, the IceCube-Upgrade~\cite{Ishihara:2019aao, TheIceCube-Gen2:2016cap} plans an even denser array with an updated DOM design to improve the capabilities of the low-energy neutrino physics program of IceCube. The focus of the IceCube-Upgrade will be on high statistic measurements of the neutrino oscillation parameters using the atmospheric neutrino flux.

  \subsection{Particle identification for very-low-energy IceCube events\label{subsec:concept}}
    Although the energy threshold of DeepCore and the IceCube-Upgrade is much lower than the full IceCube array, the nearest DOMs are still spaced \textasciitilde7m apart which, when compared to other water Cherenkov detectors such as Super-Kamiokande~\cite{Fukuda:1998mi}, is still very sparse. As a consequence, the number of PMTs which detect photo-electrons (PMT hits) in each event is relatively low, which is a problem since it is very difficult to perform any particle identification (PID) with just a few PMT hits. This has prompted the development of new optical sensors for the IceCube-Upgrade, such as the multi-PMT DOM (mDOM)~\cite{Classen:2019tlb} and the Dual optical sensors in an Ellipsoid Glass for Gen2 (DEgg)~\cite{Nagai:2019uaz}, so that more PMT hits can be collected per interaction~\cite{TheIceCube-Gen2:2016cap}. Currently at these energies, high level parameters such as the reconstructed track length are used as a discriminator to do some basic PID~\cite{Aartsen:2019tjl}, where the charge distributions get fed into the reconstruction algorithms.

    Alternatively, we use pulse shape information to perform PID. Such low level PMT information is currently unused for PID; however, given a sufficient deposited PE yield, where particles propagate within a few meters from the sensor, the pulse shape shares characteristics of the parent particle. More specifically, minimally ionizing particles (MIPs) tend to deposit photons in a short amount of time, whereas electromagnetic showers (EM showers) produce a photon spectrum with a wider distribution. At the energy range of interest here at a few~GeV, the particle content of MIPs corresponds to muons and pions, and EM showers comprise of electrons and high-energy photons.

    In this beam test, we utilize a tank filled with distilled water, upon which a PMT in a glass housing is floating, observing Cherenkov light inside the tank. We study and confirm if pulse shape information can be used for PID between MIPs and EM showers in the context of future water or ice Cherenkov neutrino telescope experiments.

\section{The Fermilab Test Beam Facility\label{sec:ftbf}}
  \subsection{FTBF MTest beam line\label{subsec:mtest}}
    \begin{figure}[!b]
      \centering
      \includegraphics[width=\textwidth]{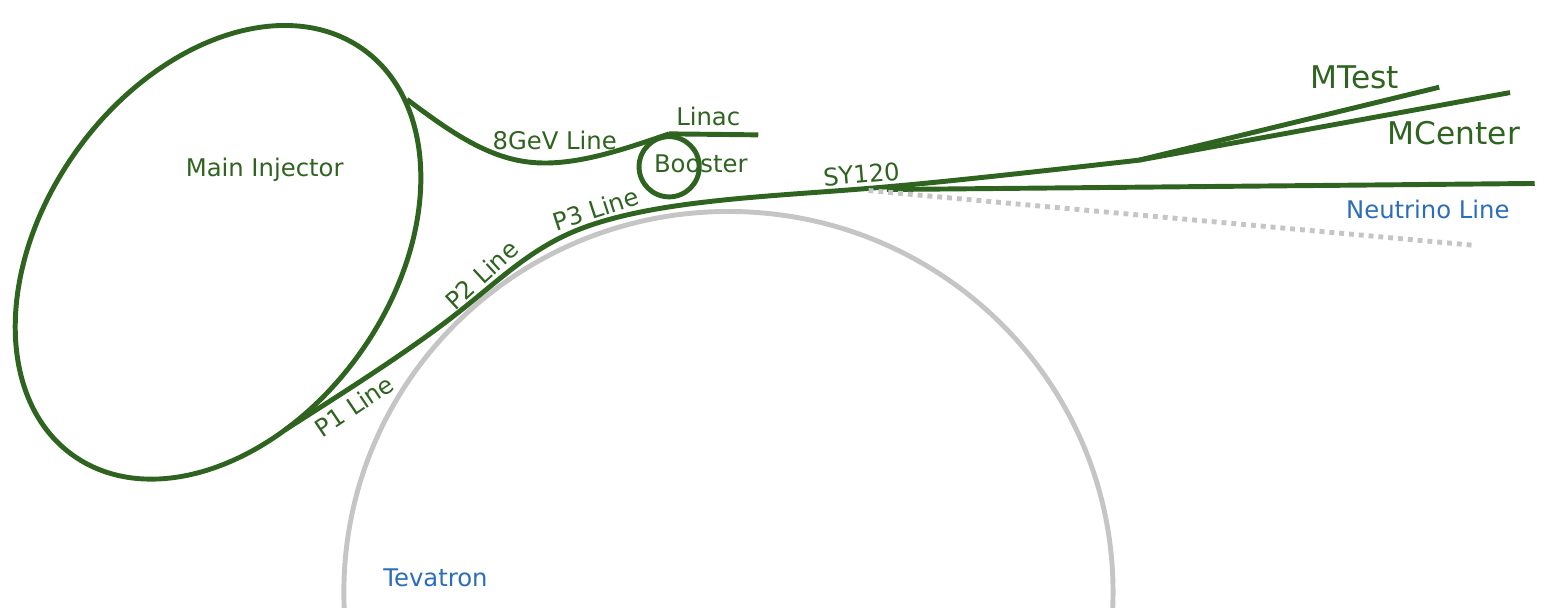}
      \caption{\label{fig:ftbf_delivery} Schematic diagram of the beam preparation of the MTest beam at the FTBF~\cite{FNAL:FTBF}. 120~GeV protons are extracted from the Main Injector and through the SY120 switchyard, they are directed to form the MTest and MCenter primary beam lines.}
    \end{figure}
    The Fermilab Test Beam Facility (FTBF) at the Fermi National Accelerator Laboratory (FNAL)~\cite{FNAL:FTBF} provides researchers with open access to high-energy and high-intensity beam lines. Two beam lines are available for use: the MTest beam line and the MCenter beam line. For the beam test discussed in this report we used the MTest beam line, as this is the beam line that is appropriate for short term experiments. At the end of this beam line, there is a wide area to place experiment-specific instrumentation. For this beam test, the beam time allocated was from 14th~June to 27th~June~2017~\cite{Rominsky:2018boq}.

  \subsection{FTBF beam structure\label{subsec:beam}}
    The primary beam which is supplied to the FTBF is comprised of protons which are extracted from the Main Injector (120~GeV proton synchrotron). This beam is structured into radio frequency (RF) buckets at a frequency of 53~MHz, with a spill duration of 4.2~seconds every minute. The intensity of the beam is tunable with a maximum intensity of $5\times10^5$~protons per spill. A schematic diagram of the primary beam preparation is shown in Fig.~\ref{fig:ftbf_delivery}. The primary beam can be collided into two movable aluminium block targets (MT1 and MT4) to create secondary beams with energies as low as \textasciitilde2~GeV, consisting of pions, muons and/or electrons. More precise tuning of the beam is available in the form of three sets of focusing magnets, two dipole magnets for selecting momenta, five trim/vernier magnets for small corrections to the beam trajectory and four collimators. In general, the uncertainty on the upstream energy of the beam is between 2-3\%, however energy loss can occur as a particle propagates through the various instrumentation~\cite{FNAL:FTBF, Aidala:2017rvg}.

%  \clearpage
  \subsection{FTBF instrumentation\label{subsec:instrumentation}}
    \begin{figure}[!b]
      \centering
      \hspace*{-50pt}
      \includegraphics[width=1.2\textwidth]{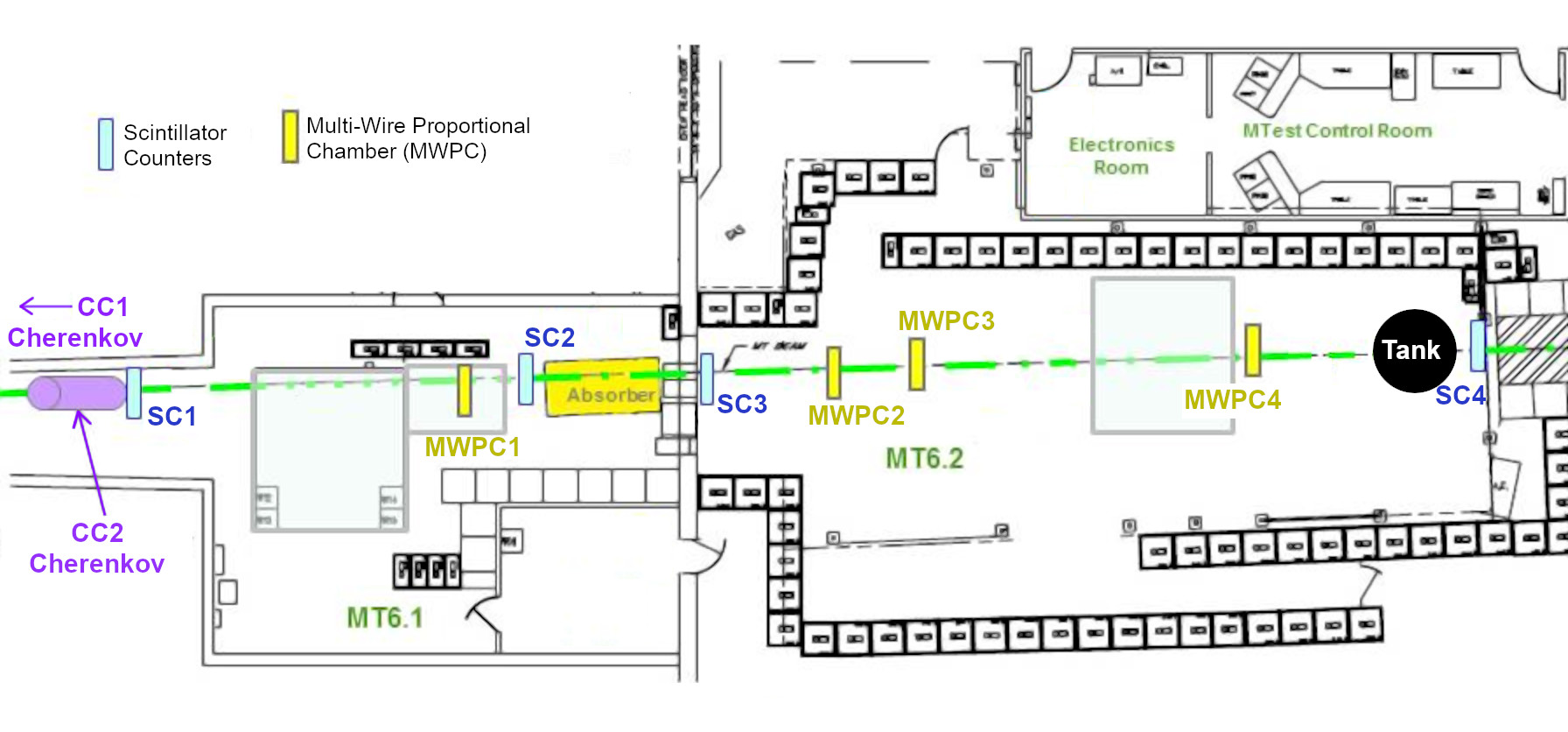}
      \caption{\label{fig:ftbf_mtest} Plan view of the MTest area highlighting the available beam instrumentation~\cite{FNAL:FTBF}. The beam enters from the left. The tank used in this beam test was placed in the location indicated by the large black circle. Two Cherenkov counters (highlighted in purple) are installed upstream for PID. Scintillators, SC1, SC2, SC3, and SC4 (shown in light blue) are placed at four positions along the beam line, with SC4 being the most downstream. MWPCs (shown in yellow) are also distributed along the beam line for beam position monitoring. NIM crates are located in the electronics room and shifters of the beam test stay in the MTest control room. These rooms are separated from the access controlled beam area (MT6.1 and MT6.2) with concrete blocks.}
    \end{figure}
\begin{figure}[p]
      \centering
      \begin{tabular}{m{0.26\textwidth}m{0.19\textwidth}m{0.47\textwidth}}
        \includegraphics[width=.27\textwidth]{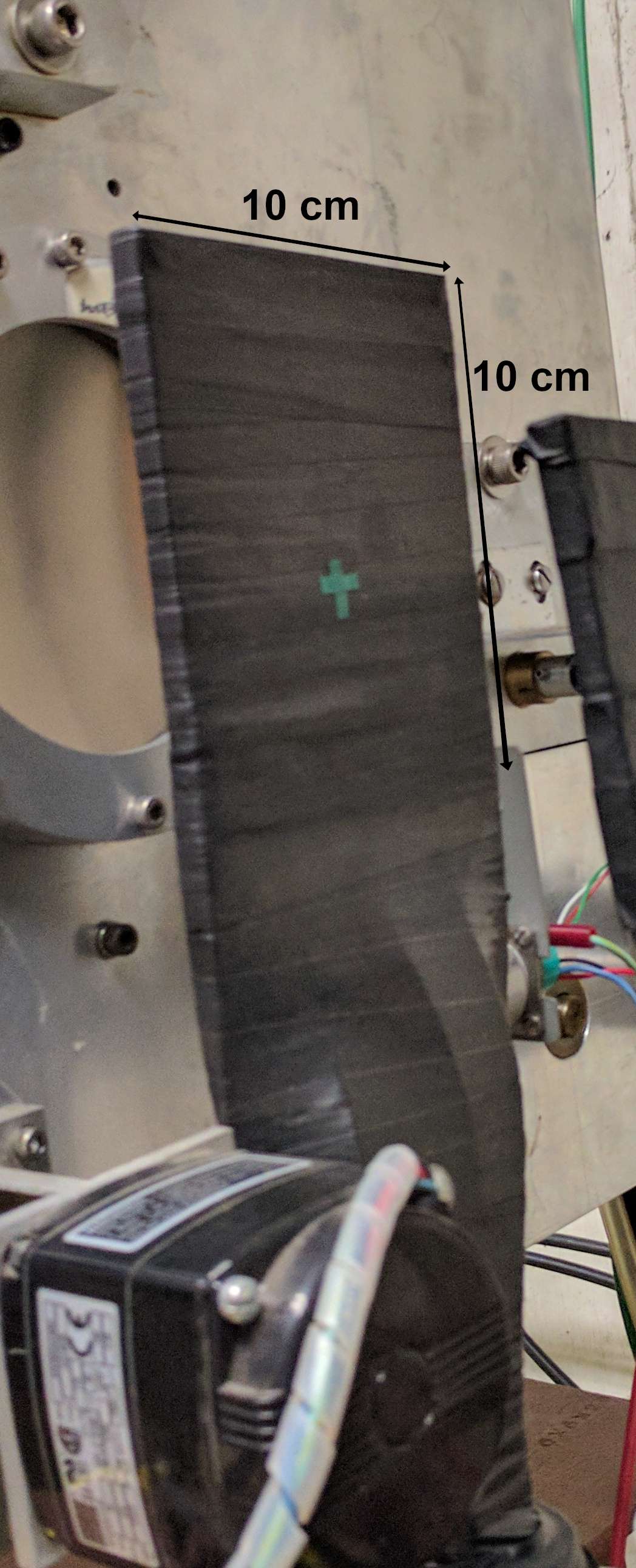} &
        \includegraphics[width=.21\textwidth]{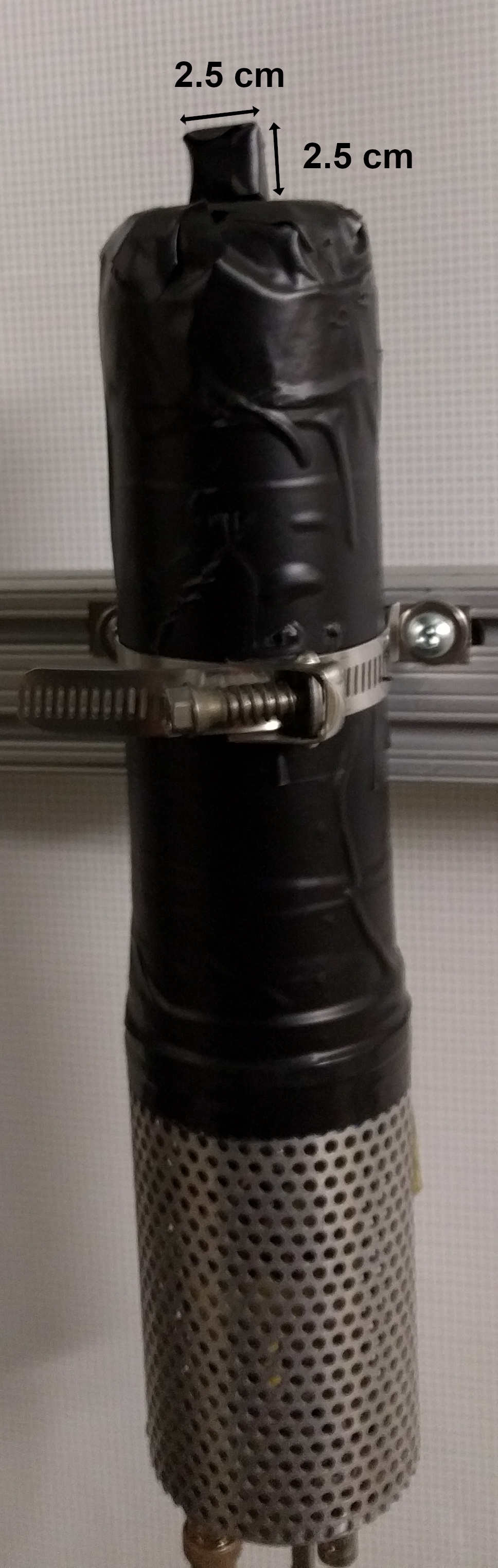} &
        \begin{tabular}{c}
          \includegraphics[width=.47\textwidth]{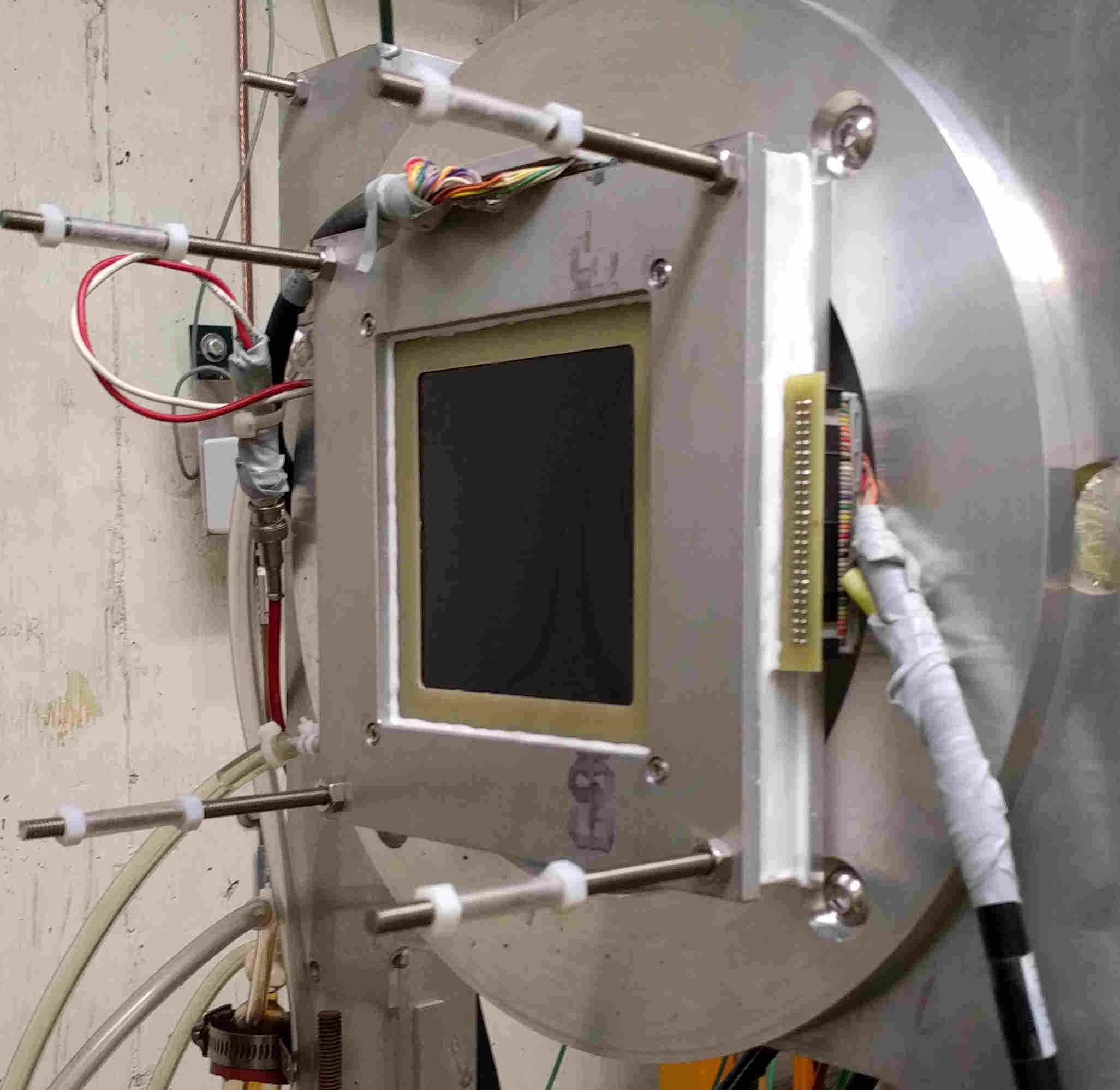} \\
          \includegraphics[width=.47\textwidth]{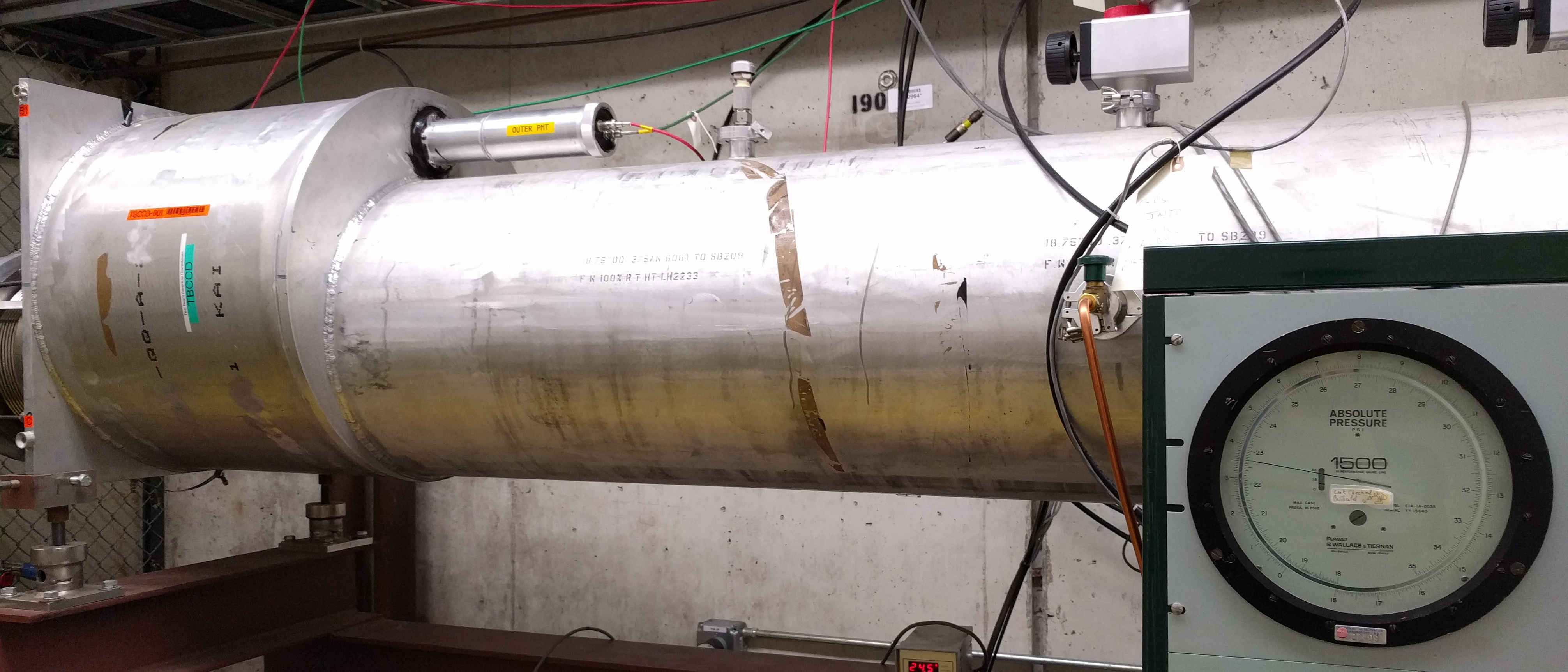}
        \end{tabular}
      \end{tabular}
      \vspace{-10pt}
      \caption{\label{fig:ftbf_ins} The FTBF instrumentation used for this beam test. The two left images show the scintillation counters: SC1 and SC4 respectively. The upper right image shows a Multi-Wire Proportional Chamber (MWPC) while the bottom right image shows a Cherenkov counter.}
      \vspace{30pt}
      \includegraphics[width=\textwidth]{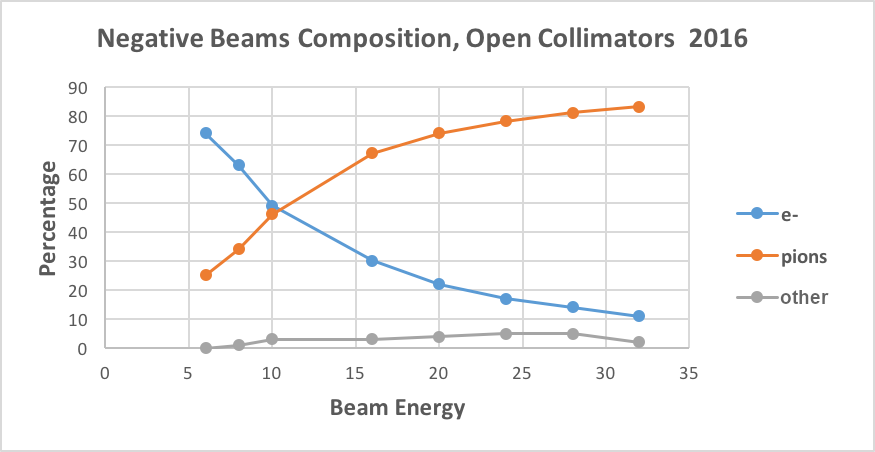}
      \caption{\label{fig:ftbf_beam_composition} Beam composition of the MTest beam line using a negative beam as a function of the beam energy in GeV
      %, as determined by E.~Skup and D.~Jensen 
      using the Cherenkov counters. The gray line (other) includes muons, kaons, and other negatively charged particles~\cite{FNAL:FTBF,Hartbrich:2016bbz}.}
    \end{figure}

    The FTBF provides multiple types of beam detector instrumentation for tracking, particle identification and triggering. A schematic of the MTest beam line is shown in Fig.~\ref{fig:ftbf_mtest}. For this beam test, we will be taking advantage of the scintillation counters, Cherenkov detectors and wire chambers. Each scintillation counter consists of a square plastic scintillator paddle connected to a PMT. There are a total of four such scintillation counters, three of them having a scintillator surface area of 10~cm~$\times$~10~cm (labelled SC1, SC2 and SC3) and one having an area of 2.5~cm~$\times$~2.5~cm (labelled SC4). There are two Cherenkov counters at MTest, one being upstream (labelled CC1) and another downstream (labelled CC2). These counters consist of 203~cm and 127~cm pressure tanks respectively, each filled with nitrogen gas as the radiative medium. These can each be utilized to enable PID based on the particle mass, by altering the gas pressure in the tanks. There are also 4 Multi-Wire Proportional Chamber (MWPC) tracking systems, located in the MTest area. Each system consists of a two-plane (X,Y) 13~cm~$\times$~13~cm wire chamber filled with argon/isobutane gas. With this, the spatial distribution and fluctuations of the MTest beam line can be profiled. Note, that the second most upstream one (MWPC2) was malfunctioning and thus was not used. Fig.~\ref{fig:ftbf_ins} illustrates the positions of the aforementioned beam instrumentation.

%\clearpage
  \subsection{Beam trigger\label{subsec:triggering}}
    For this beam test, one of the aluminium targets (MT4) was lowered to produce the secondary beam and the collimators were set to 10~mm. The focusing/dipole magnets were set to select low-energy pions and electrons of a negative polarity. A negative polarity beam was chosen in order to avoid backgrounds from protons. Fig.~\ref{fig:ftbf_beam_composition} shows the beam composition as a function of the beam energy. For this beam test, we run at energies of 4, 6 and 8~GeV with triggering set to select either pions or electrons. We also run at an energy of 2~GeV with triggering set to select electrons (note that a 2~GeV pion run was attempted, but the event rate dropped too low for any significant measurements). The intensity of the beam was kept relatively low at \textasciitilde20,000 counts on SC1 per spill. After including the coincidence condition described in the next paragraph, the maximum trigger rate seen was \textasciitilde8 triggers per spill. By taking into account the frequency of the beam, one can estimate that there is on average a \textasciitilde200~$\mu$s separation between particles incident on SC1 during a spill, leaving sufficient time between triggers. This minimizes any pileup and also keeps the trigger rate below the DAQ threshold which is around 10~Hz (see Section~\ref{subsec:daq}).

    We are interested in selecting particles which are either MIPs (muons, pions) or not MIPs, which in this beam corresponds to particles which undergo EM showering (electrons, photons). This is done using the FTBF instrumentation described in Section~\ref{subsec:instrumentation}. The coincidence of the four scintillation counters can be used to select beam particles which follow a direct trajectory into the tank. This also significantly reduces the possibility of backgrounds from cosmic rays causing contamination of our signal. Note, it was found that scintillator SC3 was misbehaving and as such it was unused, meaning that only three of the scintillation counters were used for triggering. Now that we can trigger on beam particles entering the tank, we constrain further to select MIPs, by utilizing the nitrogen gas filled Cherenkov counters. The upstream counter CC1 is set to a pressure of 200~mbar and the downstream counter CC2 is set to 240~mbar. This configuration is chosen as at the energy ranges of interest (below 10~GeV) only electrons traversing the chambers will produce Cherenkov radiation, and so by using this as a trigger we can select electrons. Note that each Cherenkov counter can individually be used to discriminate electrons, and the use of two Cherenkov counters in this beam test instead of one is purely as an extra coincidence condition taking advantage of the fact that one is upstream and the other downstream. Therefore, by using this in coincidence with the three scintillation counters, we can select electrons from the beam which penetrate the tank. The majority of the beam particles that are not electrons are pions (see Fig.~\ref{fig:ftbf_beam_composition}) therefore we can use the Cherenkov counters in anti-coincidence with the three scintillation counters to select all pions from the beam which penetrate the tank. This explains how we are able to trigger on MIPs versus EM showering particles from the beam line.

  \subsection{Beam performance\label{subsec:beam_performance}}
    \begin{figure}[t]
      \vspace*{-10pt}
      \begin{tabular}{c}
        {\includegraphics[clip,width=\columnwidth]{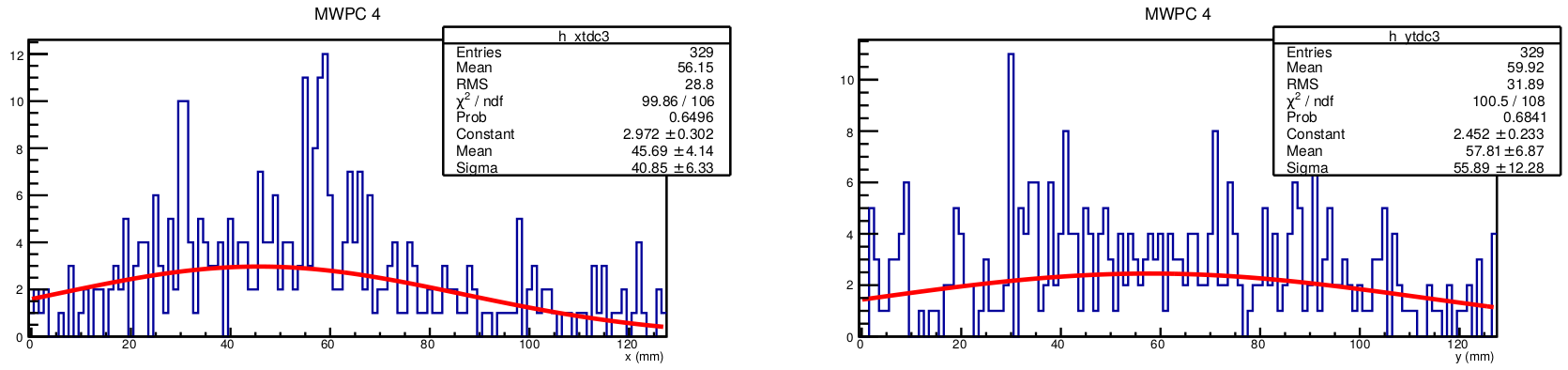}}%
        \\
        {\includegraphics[clip,width=\columnwidth]{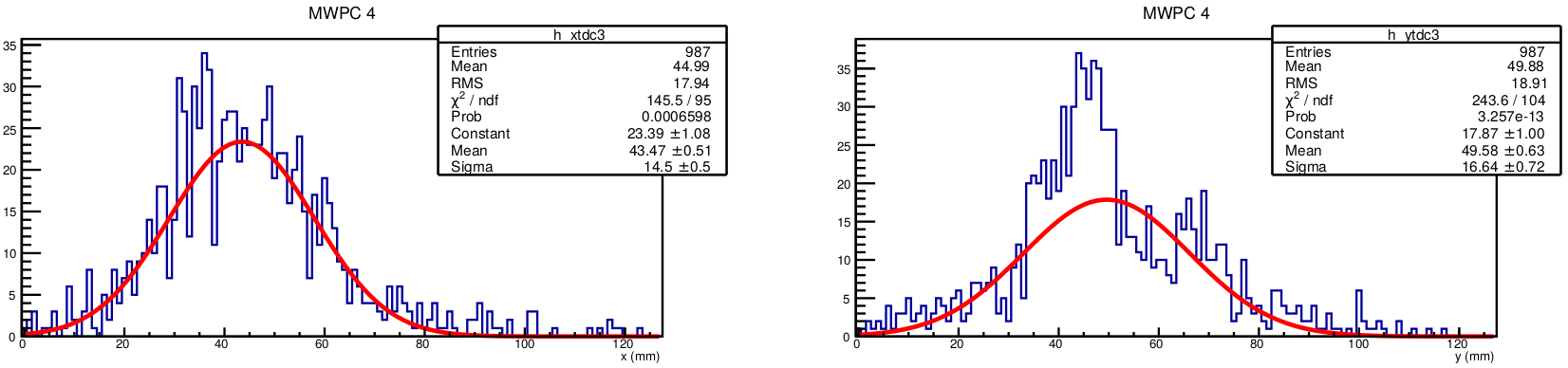}}%
      \end{tabular}
      \caption{Spatial variation of particles detected in the X-Y plane of MWPC4, which is 3~m upstream from the location of the tank. Left plots show the variation in the X-plane and right plots show the variation in the Y-plane. Upper plots show 2~GeV beam data (blue) triggering on electrons, which exhibits a broad distribution. Lower plots show 8~GeV beam data (blue) triggering on pions, which exhibits a more focused distribution with less fluctuations. It is also slightly offset from the center of the detector. A Gaussian distribution is also fitted to each of the histograms (red).}
      \label{fig:beamspread}
      \vspace*{-9pt}
    \end{figure}
    Besides the intrinsic nature of MIPs and EM showers, the beam profile can influence the shape of PMT waveforms. The PMT is susceptible to pileup which occurs when photons are recorded from more than just the triggered particle within the 320~ns trigger window. This distorts the charge distributions and pulse shape by introducing an additional signal.

    One potential source of pileup is cosmic rays. The expected rate is of order $1\times 10^{-5}$ atmospheric muons incident on the tank per RF bucket. Therefore, the likelihood of pileup occurring due to cosmic rays is extremely small.

    Another source of pileup is from the "halo" of the particle beam. This is caused by beam particles whose trajectories are outside of the three triggered scintillation counters but still enter the tank volume, while coincidentally, a particle penetrates all three scintillators and activates the trigger. If this is the case, photons from halo particles are also recorded along with the triggered particle signal.

    If the surface area of the tank volume is sufficiently covered by scintillators, the halo particles are vetoed. However, in this beam test, the area covered by the scintillators is much smaller than the area of the tank volume and they are only used to define a beam trajectory. Therefore, we need to evaluate whether or not halo particles are causing pileup. Since this is a property of the beam, we can look at the beam profile data taken with the MWPCs.

    Fig.~\ref{fig:beamspread} shows an example of data recorded from MWPC4. The upper plots show the spatial spread in X and Y for a 2~GeV beam when electrons are triggered. The lower plots show the spatial spread in X and Y for an 8~GeV beam when pions are triggered. If halo particles (which are mostly electrons) are present, they contribute to the spread of the beam. As can be seen, an 8~GeV pion beam with a pion trigger has a narrow MWPC distribution, and it was found to be roughly constant between 4 to 8~GeV. This indicates that halo particles are not present in these energy ranges and an 8~GeV beam with a pion trigger is pion dominant.

    As will be discussed in more detail later (Section~\ref{subsec:beamspread}), the 2~GeV data with an electron trigger has a wider MWPC distribution. The distribution becomes narrower at higher energies, but the 8~GeV electron beam still shows a $\sim$10~mm wider distribution than the 8~GeV pion beam (Fig.~\ref{fig:mwpc}, left). This behavior is consistent with our understanding of the beam line. Thus, whilst we expect a wide beam at 2~GeV, we cannot eliminate the possibility that halo particles are present in the 2~GeV electron data. On the other hand, the beam was run with very low rate (see Section~\ref{subsec:triggering}) and the fact that there is no indication of halo particles emerging in the 4~GeV pion beam suggests that halo particle contamination for 2~GeV data is not significant.

\section{Experimental Setup\label{sec:setup}}
  \subsection{PMT unit specification\label{subsec:dom_spec}}
    The sole photon detection and digitization unit in IceCube is the DOM~\cite{Aartsen:2016nxy}. Each module contains a Hamamatsu R7081-02 25.4~cm PMT, which has a spectral response between 300~nm to 650~nm with a peak quantum efficiency around 25\% near 390~nm. It features a box-and-line dynode chain with 10 stages, and characteristic gain and dark count being $10^7$ and $7$~kHz, respectively, at a supply voltage of $1,500$~V at 25~$^\circ$C. The DOM houses a high voltage generator, various circuit boards for digitization and calibration all inside a 35~cm diameter pressure-resistant borosilicate glass sphere made by Kopp Glass~\cite{Aartsen:2016nxy}. This particular glass material is chosen as it has a wide transparency window, down to 350~nm. Analog and digital signal processing and calibration electronics are integrated onto the mainboard and the LED flasher board. The PMT and surrounding electronics are secured in a high-strength silicon gel that optically couples components to the glass sphere.
    %A diagram of the DOM with the main components labelled is shown in Fig.~\ref{fig:dom}.
    
    The IceCube PMT used in this beam test\footnote{The PMT used here was from the DOM named "Wintery Mix".} came from a DOM in which all boards except the base board had been removed, allowing direct access to the signal and high voltage (HV) cables. This allowed us to use our own Data Acquisition (DAQ) system, instead of the traditional but bulky DOMHub DAQ~\cite{Aartsen:2016nxy}.
    From a previous lab calibration, this PMT was measured to have a gain of $3.4\times10^7$ at the operating voltage of 1500~V.

    Before moving to the full scale tank, a smaller scale setup was constructed using a standard 200~liter drum. The PMT was placed through a foam ring for stability and buoyancy, ensuring that the underside active PMT region was not masked by the foam. The drum was filled with distilled water and the PMT was floated on top. The drum was closed and wrapped in black plastic bags to prevent too many photons from leaking in. The PMT was then switched on with a voltage of 1500~V, with the signal cable connected to an oscilloscope. The trigger used here was on the leaked photon signal. This small scale setup served to verify that the PMT was functioning as expected.

  \subsection{Tank specification\label{subsec:tank}}
    \begin{figure}[!t]
      \centering
      \includegraphics[width=0.7\textwidth]{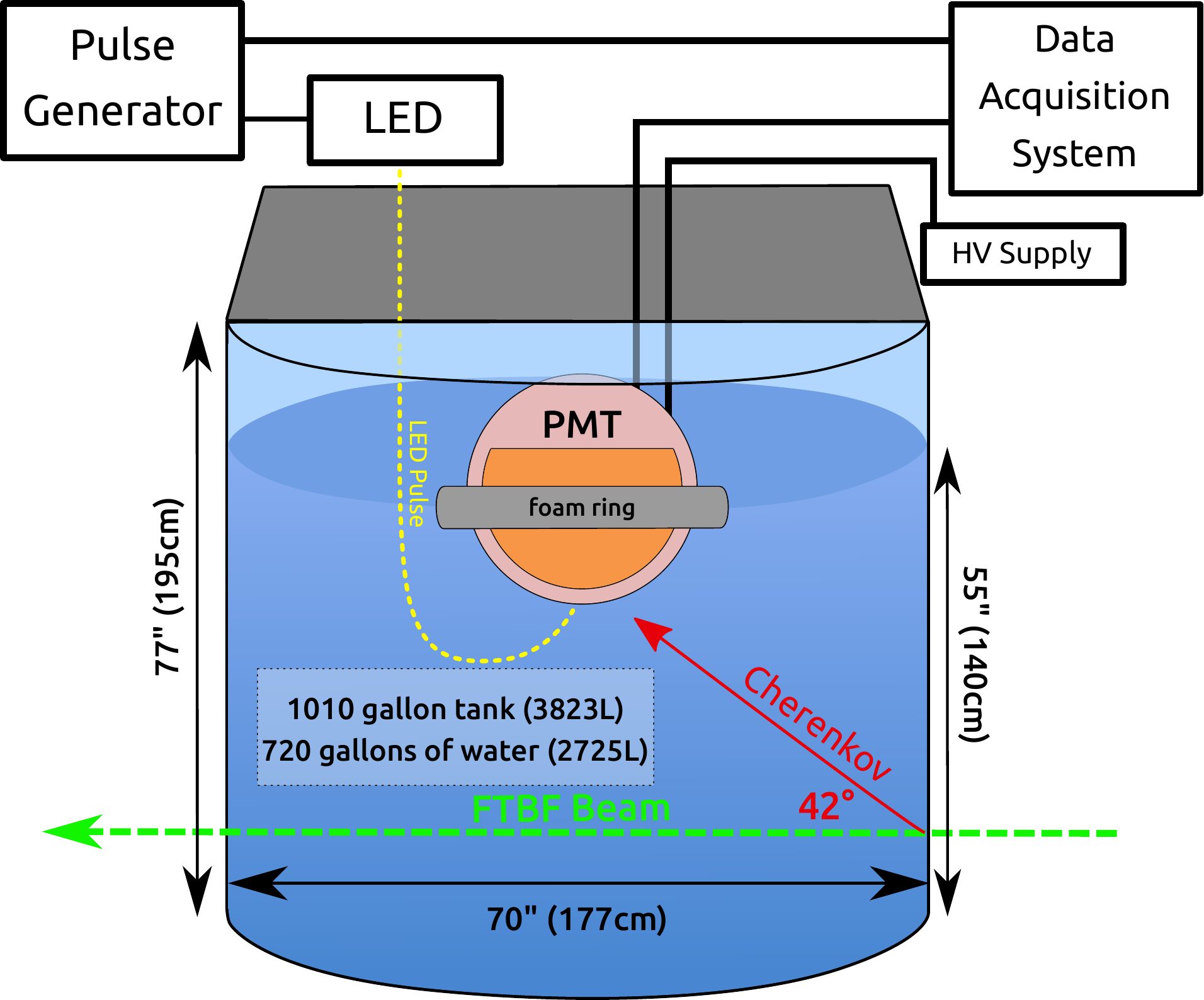}
      \caption{\label{fig:beamtest}Diagram showing the layout of the tank with the PMT placed inside. DAQ and calibration components are labelled. The LED is connected to an optical fiber thread passed into the tank to the underside of the PMT. The beam is shown by the green arrow and the propagation of the Cherenkov photons is represented by the red arrow.}
    \end{figure}
\begin{figure}[p]
      \centering
      \begin{tabular}{m{0.5\textwidth}m{0.5\textwidth}}
        \hspace{-30pt}
        \includegraphics[width=.57\textwidth]{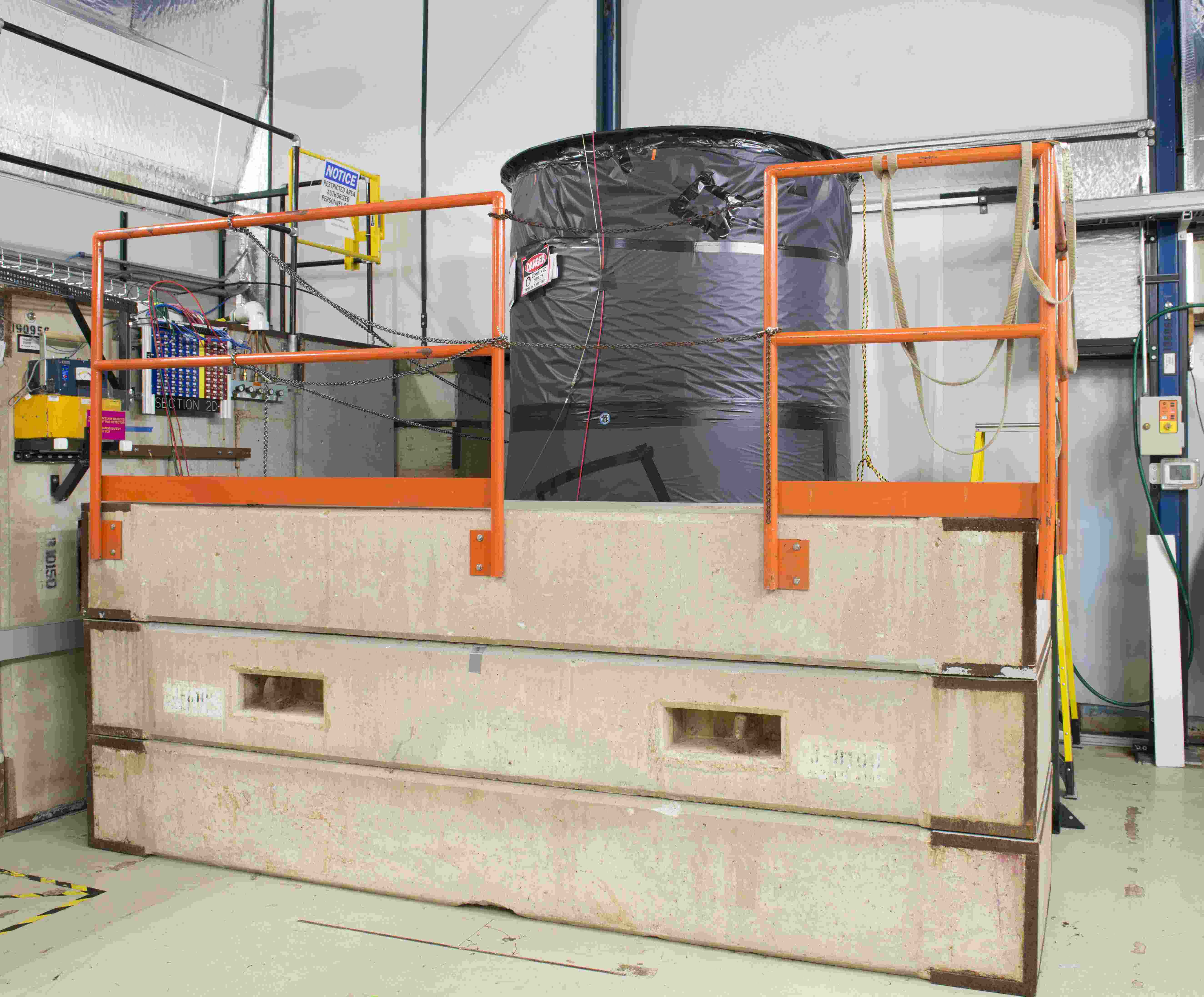} &
        \includegraphics[width=.57\textwidth]{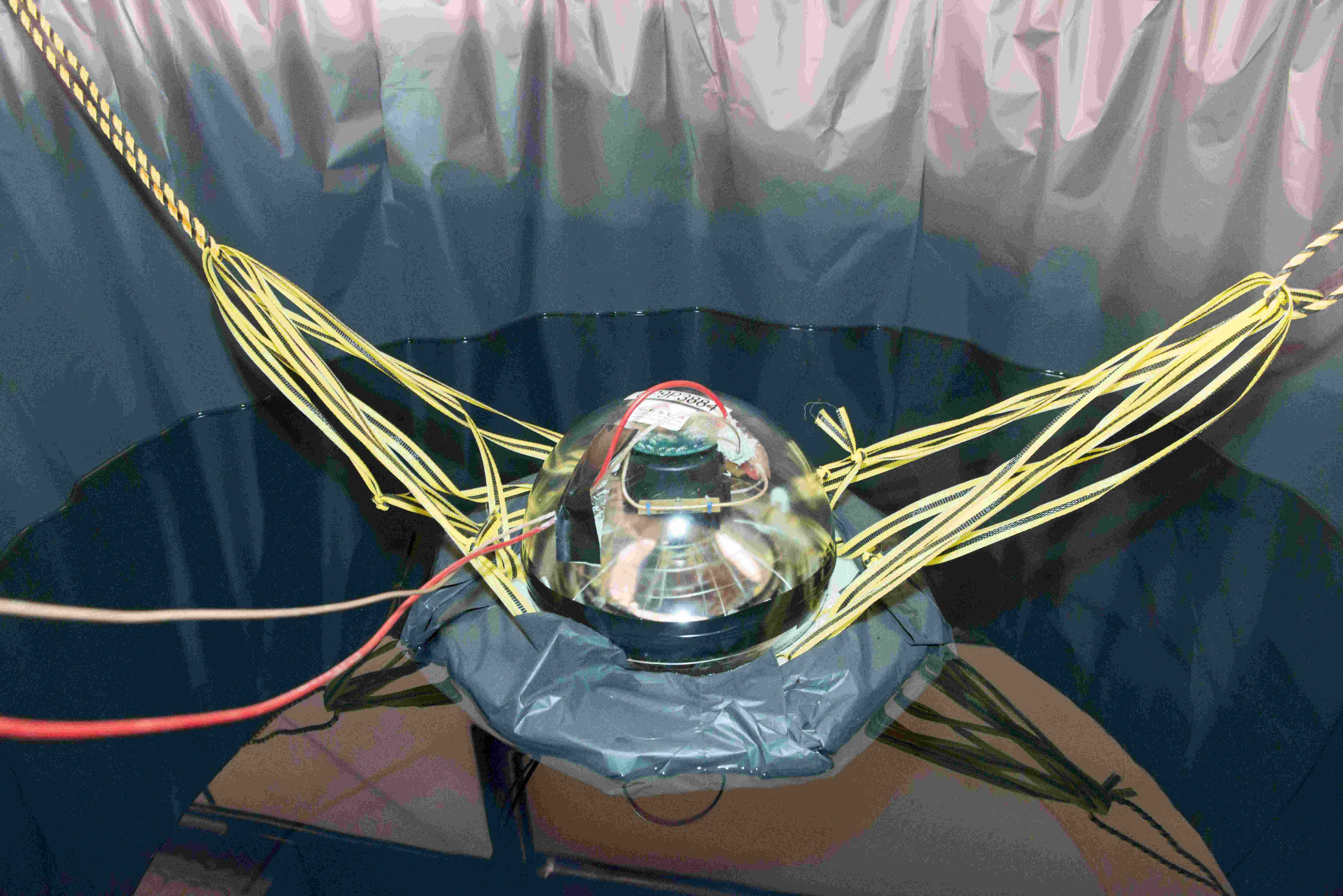}
      \end{tabular}
      \vspace{-10pt}
      \caption{\label{fig:tank} Tank placed inside the MT6.2 enclosure (left) and picture of the PMT floating inside the tank which is filled with distilled water (right).}
      \vspace{40pt}
      \includegraphics[width=0.8\textwidth]{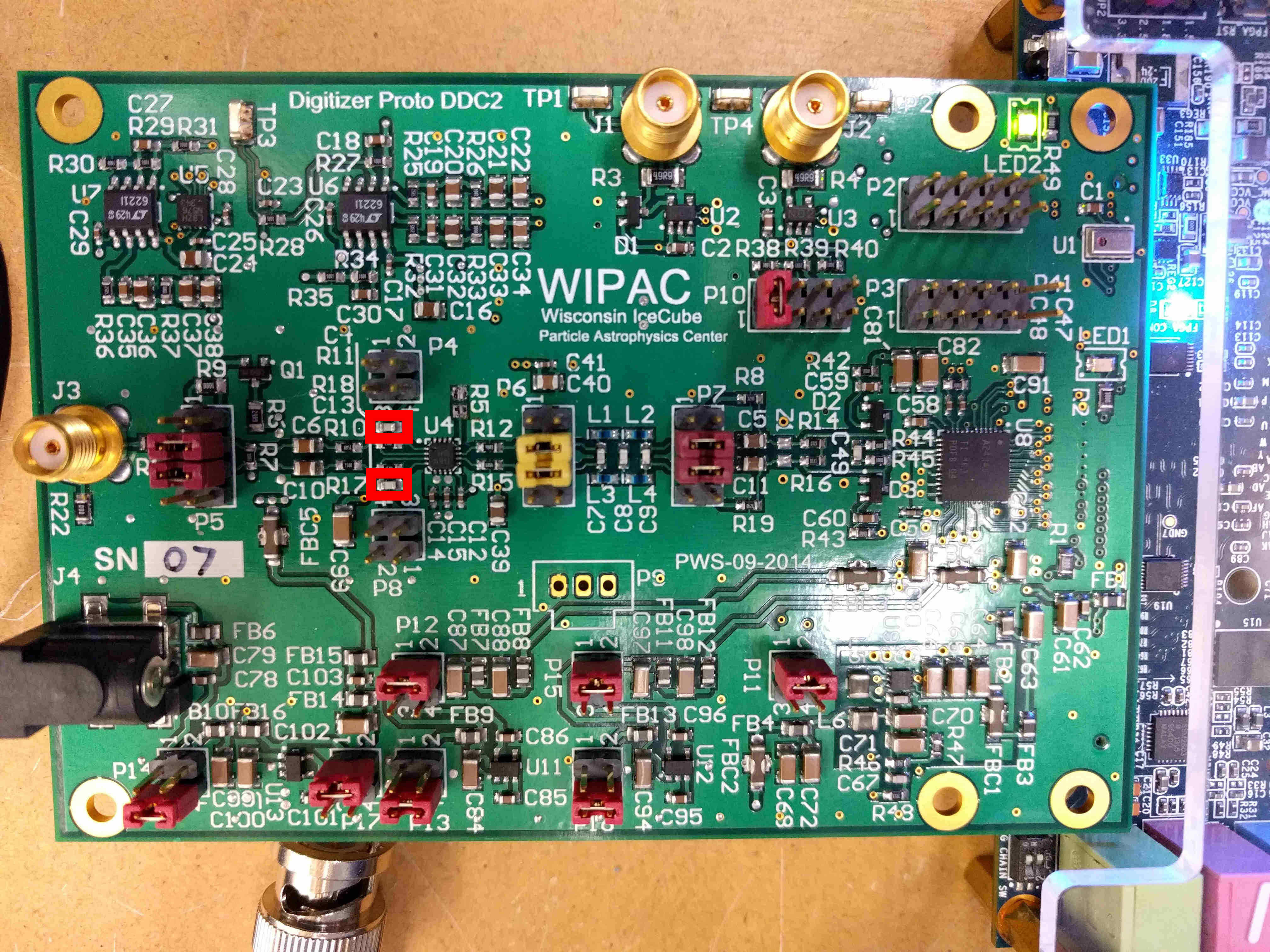}
      \caption{\label{fig:ddc2} The DDC2 (Digitizer Daughter Card, revision 2) which is the waveform digitizer used in this beam test. Highlighted in bright red are 2 capacitors, C4 and C13, which normally form the low pass filter with resistors R11 and R18. These two capacitors were removed to disable the low pass filter.}
    \end{figure}

    The tank is placed in the MT6.2 enclosure, as shown in Figs.~\ref{fig:ftbf_mtest}~\&~\ref{fig:tank}. The detector volume is contained within a cylindrical 177~$\times$~195~cm food-grade polyethylene tank, which is filled with \textasciitilde2725 litres of distilled water~\cite{TankDepot} (Fig.~\ref{fig:beamtest}). This amount of water was chosen such that the angle between the position of the PMT and the beam entry location was at 42~degrees, which corresponds to the Cherenkov angle in water. The inner and outer surfaces of the tank were coated with black Tedlar film (polyvinyl fluoride) to suppress the reflection of photons from within the tank and to reduce photon contamination from outside the tank. The PMT floated on the surface of the water with the photocathode facing down. A foam ring fitted to the PMT maintained its location (Fig.~\ref{fig:tank}, right). The opening in the ring defined the photo-sensitive area exposed to the Cherenkov radiation in the water. The beam penetrated the tank horizontally, creating Cherenkov photons which are detected by the PMT, similar to the way in which the IceCube detector operates. A commercial green LED, located outside the tank, was coupled to an optical fiber, entering the tank and pointing toward the PMT photocathode. This was used to calibrate the PMT (Section~\ref{subsec:calibration}).

  \subsection{DAQ specification\label{subsec:daq}}
    As mentioned in Section~\ref{subsec:dom_spec}, the DOM used here has had the mainboard removed, so we access the PMT signal directly. The raw PMT signal is not digitized, therefore we use a prototype Data Acquisition (DAQ) system which is being developed for IceCube Gen-2. The central element to this DAQ is the waveform digitizer DDC2 (Digitizer Daughter Card, revision 2)~\cite{Duvernois:2015pfc, Sandstrom:2014mra}. A photograph of this card is shown in Fig.~\ref{fig:ddc2}. The PMT signal is sent to the DDC2 and processed by a continuously sampling Analog-to-Digital Converter (ADC)~\cite{Abbasi:2008aa}. This converts an analog input voltage into a 14-bit digital value (an ADC count) at a rate of 4~ns/sample. At sampling frequencies above 10~MHz, the DDC2 has an AC roll-off, meaning that the ADC counts to voltage conversion (see Section~\ref{subsec:calibration}) becomes non-linear and so difficult to describe. This is in part due to a low pass filter installed in the DDC2 which filters the higher frequency signal components to make the waveform signal smoother. Since we are interested in doing a pulse shape analysis, in order to preserve the shape of the waveform, we removed the low pass filter in the DDC2, extending the AC roll-off to above 10~MHz. During data taking, we send a TTL (Transistor-Transistor Logic) signal generated from the beam monitor coincidence (Section~\ref{subsec:triggering}) as an external trigger to the DDC2. Note; since the IceCube neutrino telescope operates continuously on all signals, the ability of the DDC2 to accept an external trigger is designed-in only for test purposes.

     Prior to the beam test, the calibration of the DDC2 was performed at the University of Wisconsin-Madison. The calibration constant of the DDC2 is called the least significant bit (LSB) and this is obtained by triggering the digitizer to record a known voltage which is generated using a pulse generator. A 1~MHz sine wave of amplitude 1~V is used as the signal. From this, the LSB was found to be 0.220~mV/ADC count. Sine waves up to a frequency of 10~MHz were tested and found to be consistent with this value. After subtracting the baseline, we find the dynamic range of the DDC2 to be 2~V.

    Another element of the DAQ system is the FPGA (field programmable gate array), which is used to programme the DDC2 functions and also to interface it with a computer so that the data can be saved. The FPGA used in this beam test is the Intel (formerly Altera) Cyclone V SX FPGA~\cite{Intel:FPGA}. By connecting the DDC2 to a computer via this FPGA, the configuration settings on the DDC2 such as the triggering can be managed. During operation, when a PMT signal is produced, the digitized PMT signal from the DDC2 is handled by the FPGA and then forwarded to the computer through a USB cable as an ASCII (text) table, with an entry for the timing and also the ADC count for each sample. Transfer of data using ASCII is slow, and for this beam test we found that a triggering frequency of above \textasciitilde10~Hz starts to cause loss of data in the FPGA output. Our trigger rate is at maximum 2~Hz due to the requested low intensity beam and low coincidence rate (see Section~\ref{subsec:triggering}), therefore this limitation does not cause any issues. It would be more efficient to use a binary format for data transfer and this is currently in development.

  \subsection{Calibration\label{subsec:calibration}}
    \begin{figure}[p]
      \centering
      \includegraphics[width=0.8\textwidth]{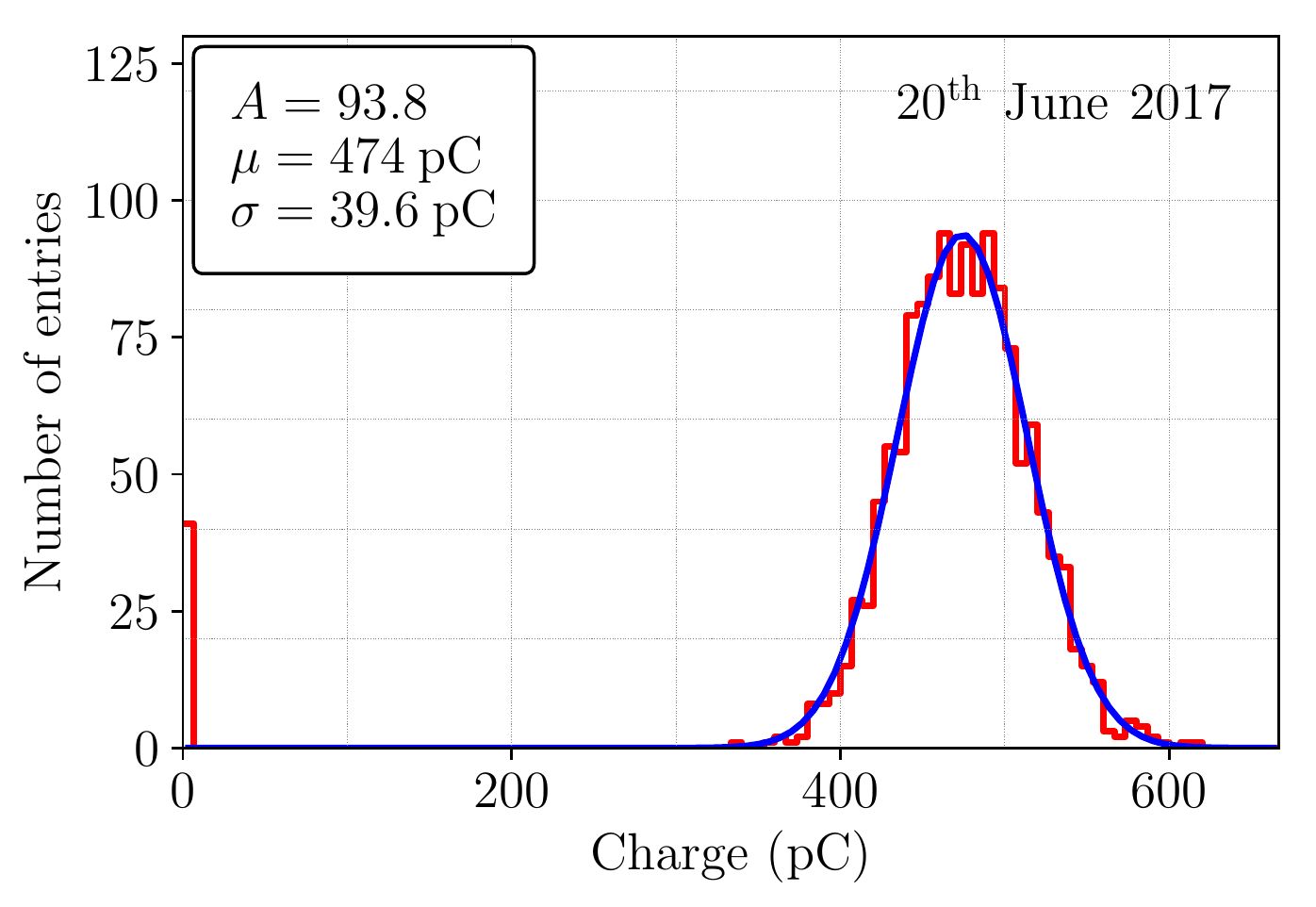}
      \caption{\label{fig:led_charge} Charge distribution of data collected with the PMT on 20$^{\rm th}$ June 2017 of a flashing LED pulse. Fitted to this is a Gaussian distribution with normalization $A$, mean $\mu$ and standard deviation $\sigma$.}
      \vspace{30pt}
      \begin{tabular}{m{0.5\textwidth}m{0.5\textwidth}}
        \hspace*{-35pt}
        \includegraphics[width=.55\textwidth]{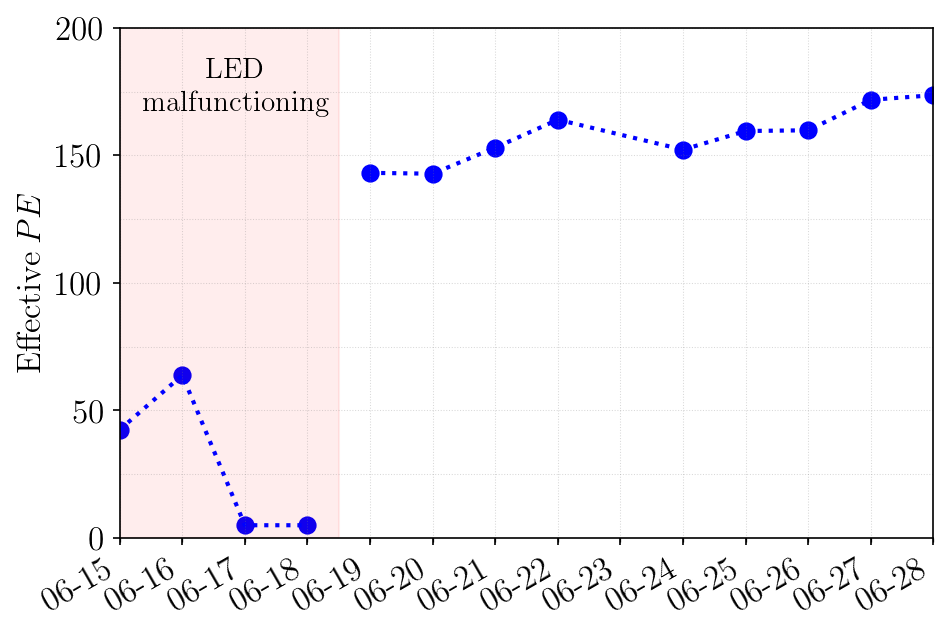} &
        \hspace*{-20pt}
        \includegraphics[width=.55\textwidth]{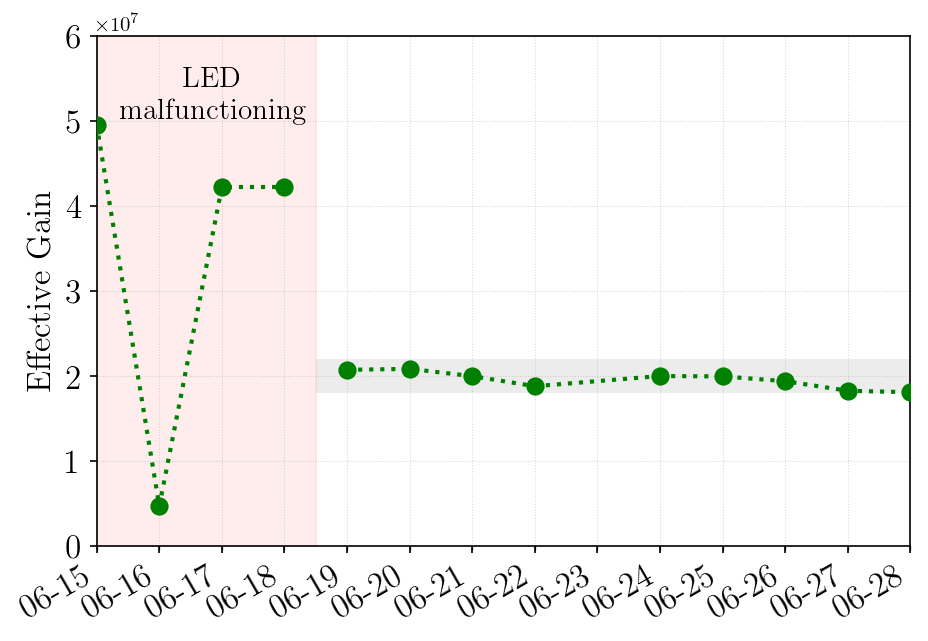}
      \end{tabular}
      \vspace{-10pt}
      \caption{\label{fig:gain_pes} On the left, a plot showing the variability of the average number of effective PE liberated from the photocathode per waveform as a function of the date and on the right a plot showing the variability of the effective gain as a function of the date, with a 10\% error shaded in gray. A typical LED waveform is 200~ns wide. For the dates June 15--June 18, the LED was found to be malfunctioning.}
    \end{figure}

    The gain variation of the PMT is monitored daily by the LED system which is shown in Fig.~\ref{fig:beamtest}. This LED is a standard commercially available green LED. A 4~V square pulse with a width of 40~ns was used to flash the LED and also used as an external trigger on the DDC2, to record waveforms detected from the LED photons. Each of the recorded waveforms is integrated to get the total charge deposited on the PMT. A charge histogram is produced, such as the one shown in Fig.~\ref{fig:led_charge}. A Gaussian distribution with a normalization $A$, mean $\mu$ and standard deviation $\sigma$ is fitted to the distribution. As outlined in Appendix~\ref{sec:gain_calculation}, by assuming the charge distribution is proportional to the true photon statistics, the average number of PE liberated from the photocathode per waveform ($PE$) and the PMT gain ($g$) can be estimated using the following equations:
    \begin{align}
      PE&=\left(\frac{\mu}{\sigma}\right )^2 \label{eqn:PE}\\
      g&=\frac{\mu}{PE\cdot C\cdot R}
    \end{align}
    where $g$ is the gain, $C$ is the charge of a single electron i.e.\ $1.6\times10^{-19}$~C and $R$ is the impedance, which for the DDC2 is 150~$\Omega$.

    Fig.~\ref{fig:gain_pes} shows the stability of the gain and PE as a function of the day. We later found that the LED intensity saturates the PMT which breaks the relation shown in Eq.~\ref{eqn:PE}, therefore the extracted PE and gain are only effective. Thus in this analysis, we rely on the known gain of this PMT at 1500~V. Effective PE and gain can still, however, demonstrate the stability of both the PMT and the transparency of the water as a function of time. The LED was seen to be malfunctioning in the first 4 days of data taking, and so on June~19, we visually inspected the LED and found it to be very wet. The LED was then moved further from the tank to prevent water condensation. After this, the effective gain and PE can be seen to be stable with time, the effective gain being at a value of $2\times10^7$ and the effective PE at \textasciitilde150. This tells us that the experimental setup was stable over the period of our data taking. From this measurement we assign a 10\% error on the PMT gain (as shown on the figure). By assuming the stability of the LED system, we can conclude that water transparency degradation is not a problem for any of our measurements. 

\vspace*{-4pt}
\section{Results\label{sec:results}}
\vspace*{-4pt}
  \begin{figure}[p]
    \centering
    \hspace*{-40pt}
    \includegraphics[width=1.2\textwidth]{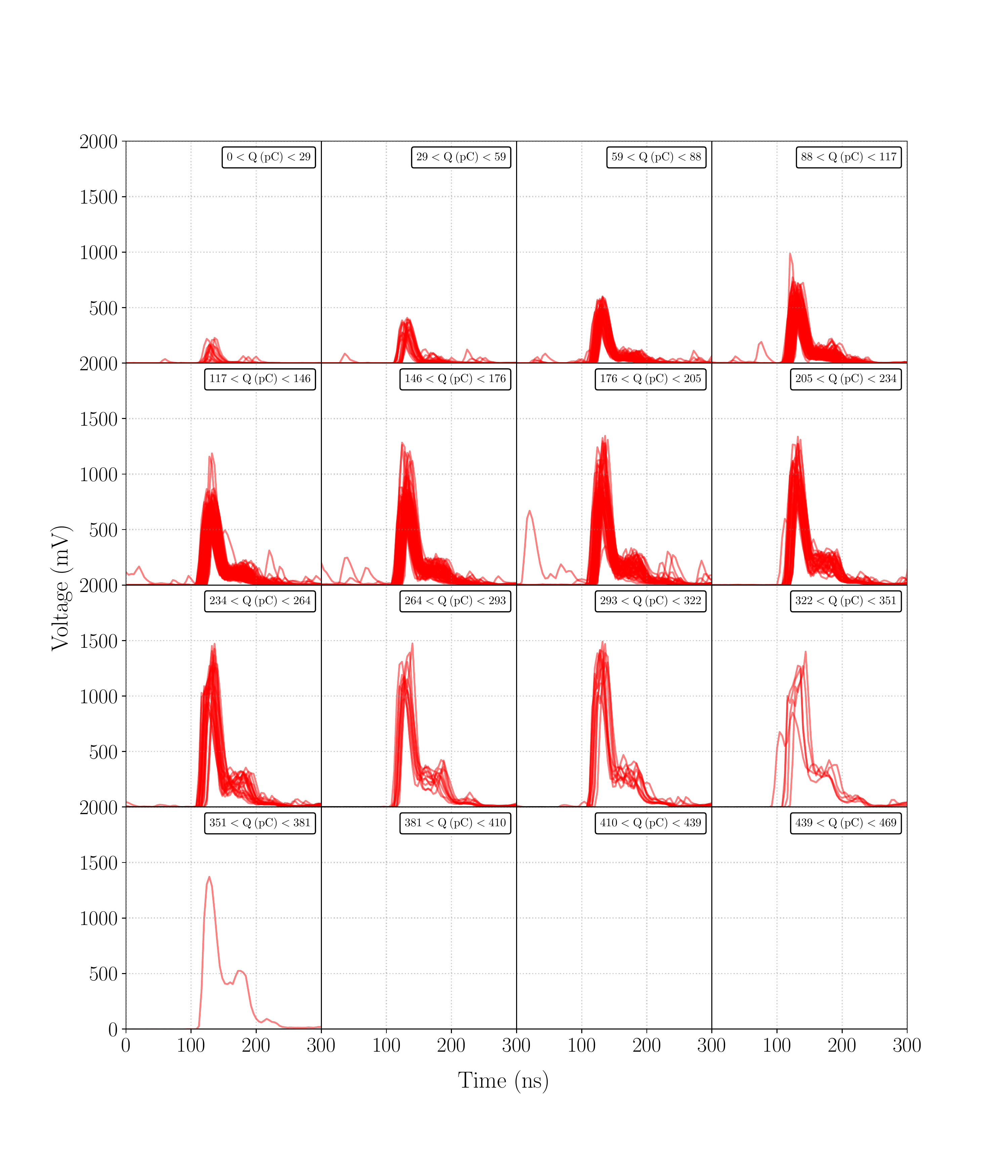}
    \vspace*{-40pt}
    \caption{\label{fig:waveform-e} PMT waveforms for a 2~GeV electron beam, split up into respective charge bins labelled on the top right of each plot.}
  \end{figure}
  \begin{figure}[p]
    \centering
    \hspace*{-40pt}
    \includegraphics[width=1.2\textwidth]{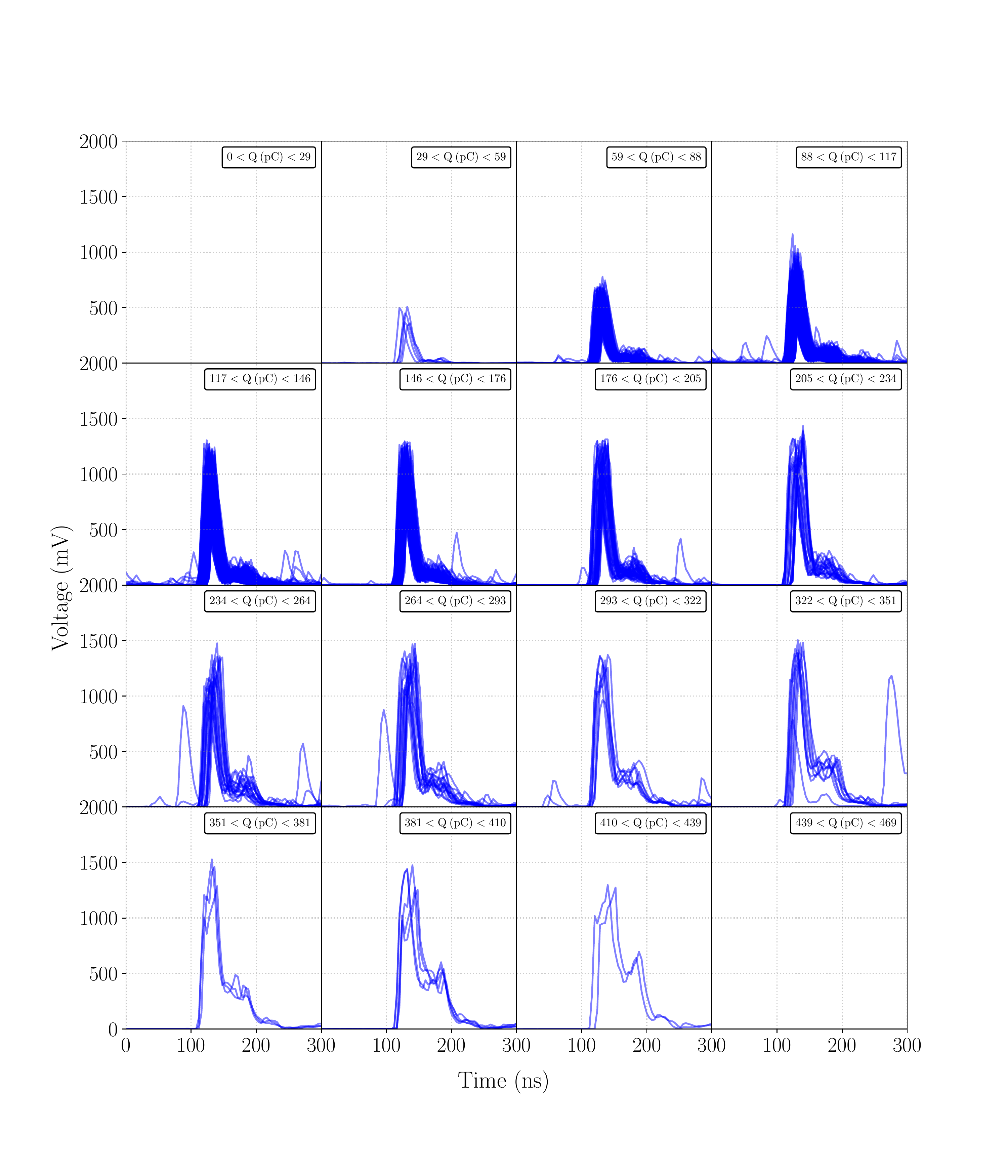}
    \vspace*{-40pt}
    \caption{\label{fig:waveform-pi} PMT waveforms for an 8~GeV pion beam, split up into respective charge bins labelled on the top right of each plot.}
  \end{figure}

  \subsection{PMT waveforms\label{subsec:pmt_waveforms}}
  Figs.~\ref{fig:waveform-e}~\&~\ref{fig:waveform-pi} show the digitized waveforms for 2~GeV electron and 8~GeV pion data. We focus on performing the pulse shape analysis on these data sets as at these energies an EM shower and a MIP like particle deposit similar charges. Each digitized waveform represents a triggered event that produced Cherenkov radiation in the detector volume. Each triggered event has a recording window of 320~ns. 100~ns of this is pre-trigger, therefore we record up to \textasciitilde200~ns after the primary pulse. Waveforms are plotted into 16 charge bins.

    Each charge bin contains a collection of waveforms. From an initial observation, both electron and pion produced waveforms appear to have a marginal difference in width and share a reasonably similar shape. Each waveform consists of a primary and secondary pulse located between 100--200~ns for each waveform. The amplitude of the first pulse in each bin grows with charge until \textasciitilde1500~mV, which may be an indication of saturation (see Section~\ref{subsec:saturation}). Similarly, the secondary pulse demonstrates a linear growth with increasing charge.

    At this stage, it is difficult to distinguish between the electron and pion produced waveforms. Further pulse shape analysis is explored by:
    \begin{itemize}
      \itemsep0em
      \item \vspace{-3pt}Characterizing the primary pulse.
      \item \vspace{-5pt}Characterizing the secondary pulse.
      \item \vspace{-5pt}Understanding saturation of the pulse amplitude despite the sufficient dynamic range.
      \item \vspace{-5pt}Determining whether the spread of the beam influences the width of the pulses.
    \end{itemize}

  \subsection{Primary pulse\label{subsec:primary_pulse}}
  \begin{figure}[p]
    \centering
    \hspace*{-40pt}
    \includegraphics[width=1.2\textwidth]{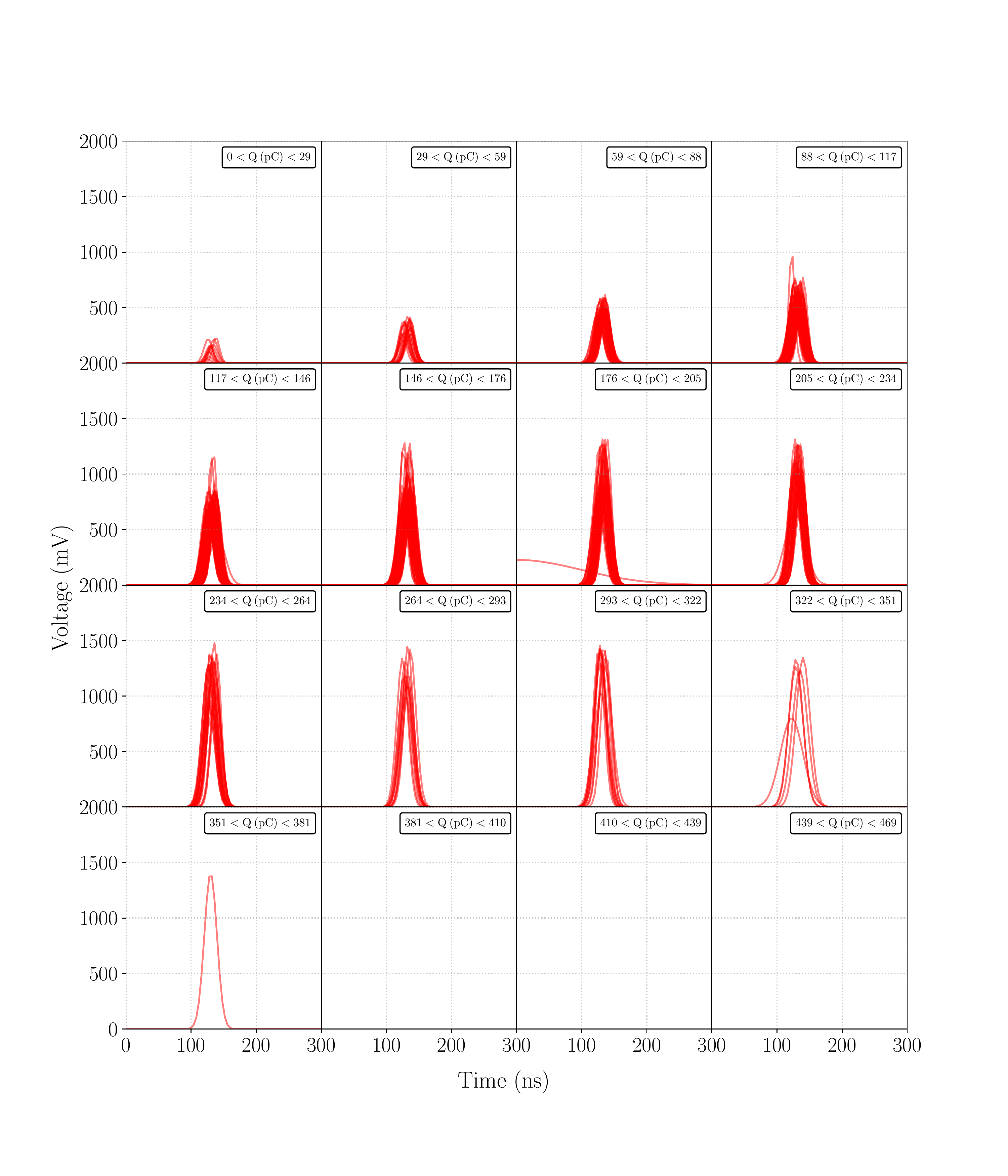}
    \vspace*{-40pt}
    \caption{\label{fig:waveform-e-gauss} Primary pulse extracted with a Gaussian fit for a 2~GeV electron beam, split up into respective charge bins labelled on the top right of each plot.}
  \end{figure}
  \begin{figure}[p]
    \centering
    \hspace*{-40pt}
    \includegraphics[width=1.2\textwidth]{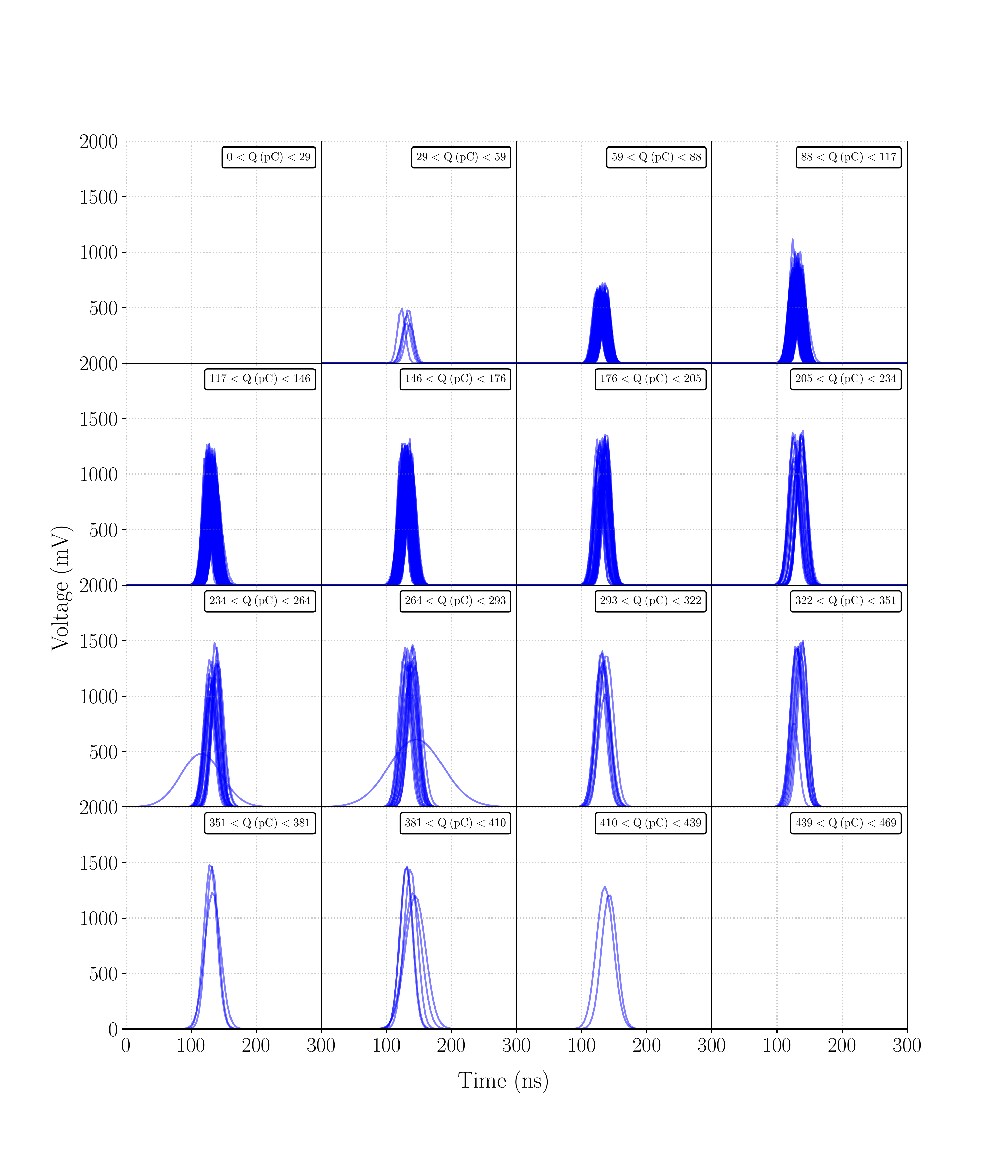}
    \vspace*{-40pt}
    \caption{\label{fig:waveform-pi-gauss} Primary pulse extracted with a Gaussian fit for an 8~GeV pion beam, split up into respective charge bins labelled on the top right of each plot.}
  \end{figure}
    In order to identify the features of the waveform, we split the waveform into two regions: a primary pulse and a secondary pulse. The primary pulse occurs in the timing region of \textasciitilde100-150~ns and it reflects the pulse generated by the PMT from standard photo-multiplication. To isolate the primary pulse, a Gaussian distribution with normalization $A_{p}$, mean $\mu_{p}$, and standard deviation $\sigma_{p}$ is fitted, ignoring the waveform contributions above 150~ns so as to approximate only the primary pulse. The produced pulses can be seen in Fig.~\ref{fig:waveform-e-gauss} for a 2~GeV electron beam and Fig.~\ref{fig:waveform-pi-gauss} for an 8~GeV pion beam. For the majority of waveforms, this procedure can be seen to be a reasonable estimate of the primary pulse. However, there are outliers which either did not fit well to the waveform or simply did not fit all, either because the fit failed or because the waveform was anomalous. These can be identified and removed simply by requiring $100$~ns~$<\mu_p<150$~ns and $2.5$~ns~$<\sigma_p<15$~ns, thus only selecting the higher quality waveforms. In total, this cut removed \textasciitilde2\% of waveforms. All primary pulse results shown from now will have this selection imposed. Pulse shape analysis can now be done to compare the two datasets, but first we look into the other three points discussed above to understand the impact they might have on the pulse shape analysis.

\clearpage
  \subsection{Secondary pulse\label{subsec:second_pulse}}
    \begin{figure}[b]
      \centering
      \includegraphics[width=0.7\textwidth]{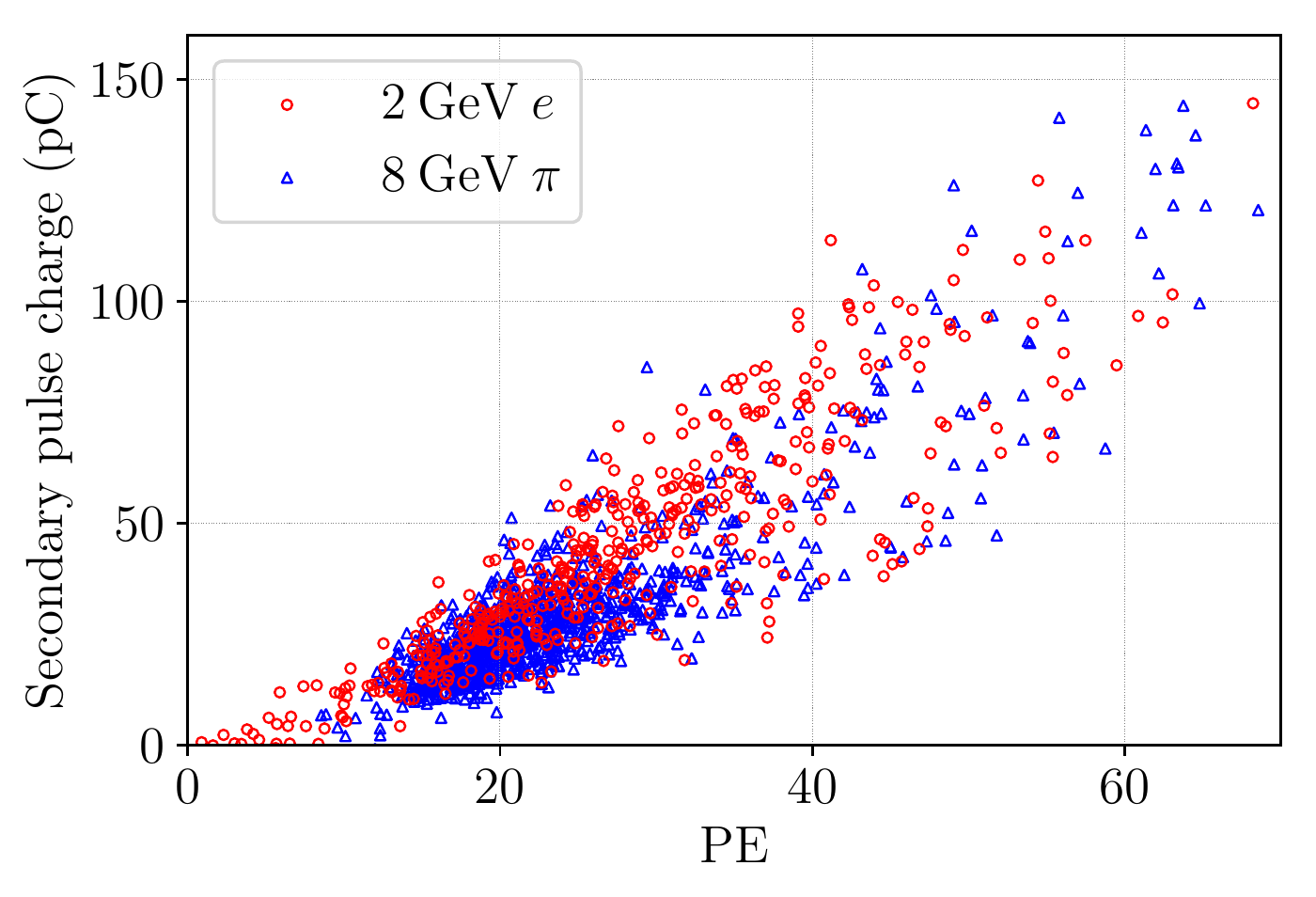}
      \caption{Charge of the secondary pulse plotted as a function of the charge of the waveform. In red circles show data points for a 2~GeV electron beam and in blue triangles show data points for an 8~GeV pion beam. For both configurations, it can be seen that the second pulse grows linearly with the total waveform.\label{fig:sp_growth}}
    \end{figure}
    The waveforms in Figs.~\ref{fig:waveform-e}~\&~\ref{fig:waveform-pi} can be seen to exhibit a non-Gaussian secondary peak that grows with increasing charge at timings of above \textasciitilde150~ns. Such behavior is a common feature of PMTs and can be described as:
    \begin{itemize}
      \itemsep0em
      \item \vspace{-4pt}Afterpulse: Electrons accelerated between dynodes induce ionization of residual gas molecules. Afterpulsing is described to grow linearly with charge; however, the timing range they are expected to be seen in is from 300~ns to 11~$\mu$s, and so the secondary pulse seen in the waveforms here are not likely due to afterpulsing~\cite{Abbasi:2010vc}.
      \item \vspace{-4pt}Late pulse: The primary PE which impacts the first dynode elastically or inelastically backscatters. It briefly decelerates and accelerates again towards the dynode chain, due to the electric field. The delay between the first and second peak should equal twice the PE transit time between the photocathode and amplification chain. The resultant peak is completely separate from the main pulse with a broadened response time. Generally, this occurs on a timescale of up to 70~ns after the primary pulse and is expected to grow linearly with increasing charge~\cite{Kaether:2012bm}.
    \end{itemize}
    From the timing, the secondary peak is likely to be due to late pulses, as we record \textasciitilde200~ns of the waveform after the primary peak. To confirm this we check whether there is a linear relationship between the total charge and the secondary peak charge. To calculate the secondary peak charge, the charge of the Gaussian primary peak is subtracted from the total charge of the waveform (see Sec~\ref{subsec:primary_pulse}). Fig.~\ref{fig:sp_growth} shows the secondary pulse charge as a function of the total charge of the waveform for both the 2~GeV electron beam and the 8~GeV pion beam. This plot demonstrates the linearity between the two, and so confirms that the secondary pulse is due to late pulses. The linear behavior continues even for high values of charge, suggesting that it is not limited by saturation effects of the primary pulse.
    
  \subsection{Saturation\label{subsec:saturation}}
    PMT saturation arises when the number of PE impacting the dynode chain exceeds its amplification capability. This is because the maximum PMT output is limited by the current flowing to the voltage divider circuit~\cite{Hamamatsu}. The onset of saturation causes the voltage read out from the PMT to become non-linear with respect to the number of incident PE until, with increasing PE, it eventually plateaus.

    Figs.~\ref{fig:waveform-e}~\&~\ref{fig:waveform-pi} show the effect of PMT saturation. The waveform voltage stops increasing around \textasciitilde1500~mV suggesting the PMT is saturated. For a given energy, waveforms for the electron data experience more saturation compared to waveforms from pion data because of the larger charge deposition by EM showers vs MIP tracks. Saturation becomes an issue for electron data taken at high energies, motivating us to use the lower energy electron data for this analysis.

  \subsection{Beam spread\label{subsec:beamspread}}
    \begin{figure}[b]
      \centering
      \begin{tabular}{m{0.5\textwidth}m{0.5\textwidth}}
        \hspace*{-35pt}
        \includegraphics[width=.55\textwidth]{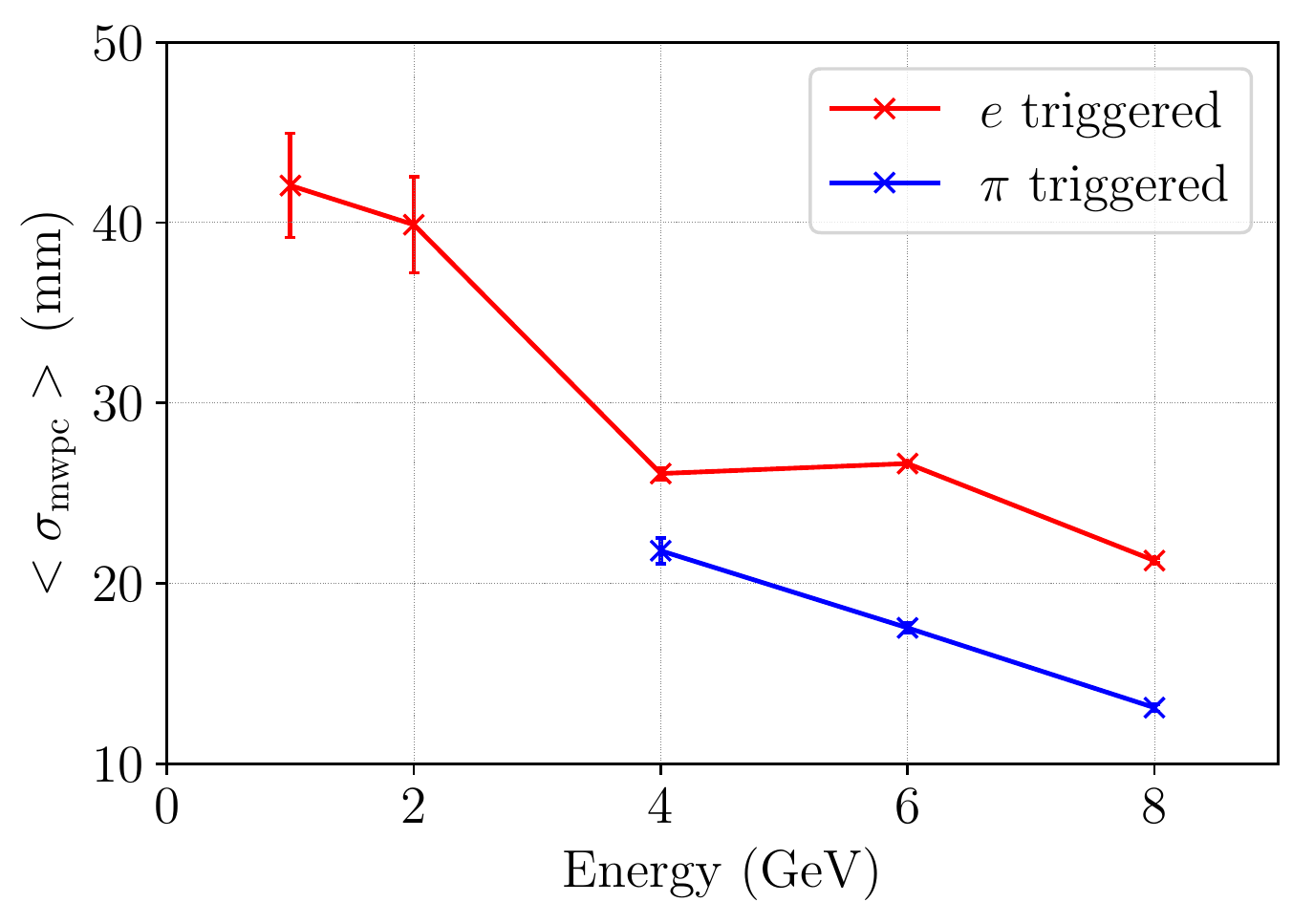} &
        \hspace*{-20pt}
        \includegraphics[width=.55\textwidth]{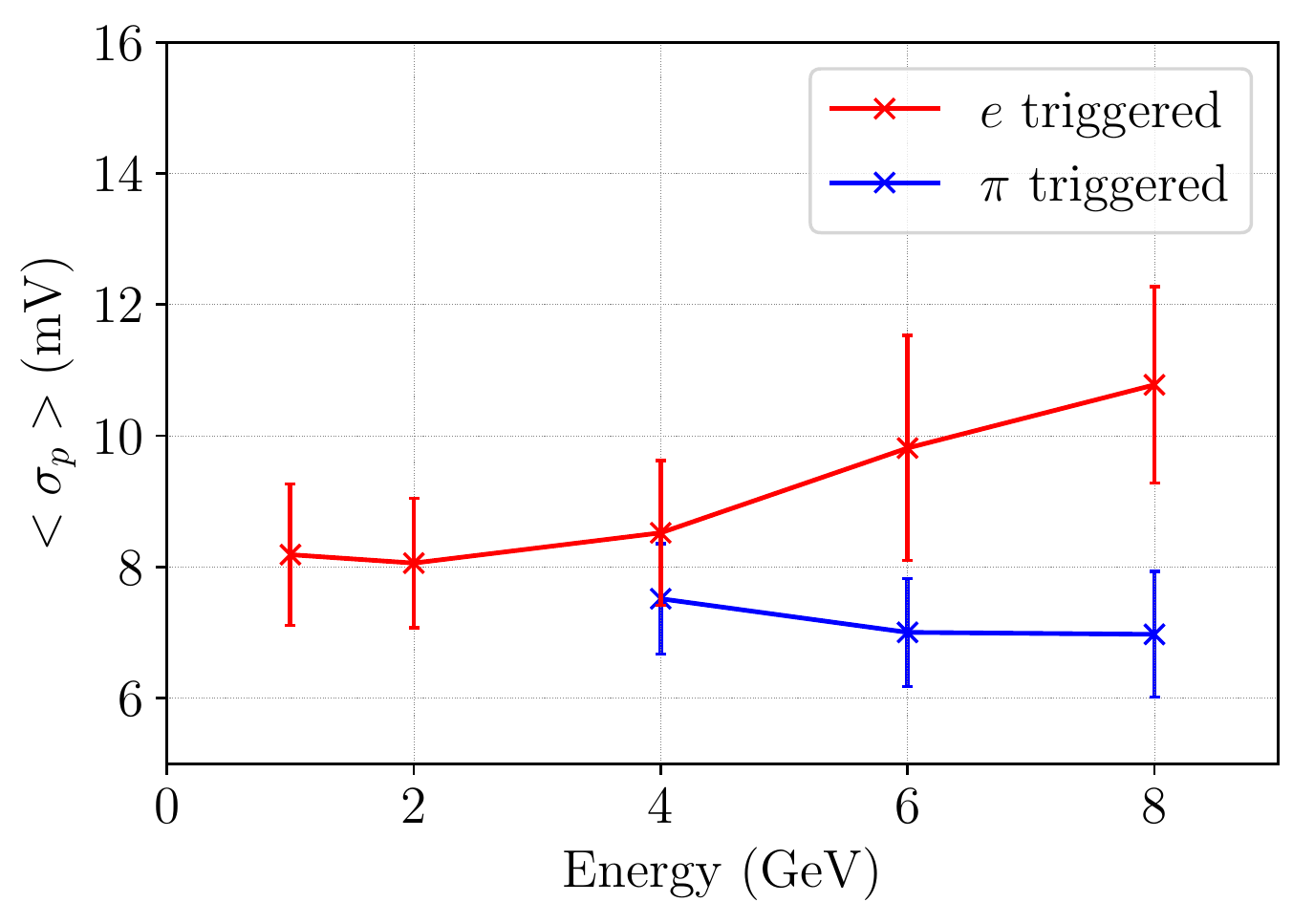}
      \end{tabular}
      \caption{Left figure shows the average spread of the beam as a function of the beam energy for an electron trigger and pion trigger. Spread was computed from beam profile data which was taken using 3 multi-wire proportional chambers (see Fig.~\ref{fig:beamspread}). The right figure shows the average of the primary peak width as a function of the beam energy. In both figures electron and pion triggered data are shown in red and blue, respectively.
      \label{fig:mwpc}}
    \end{figure}

    As mentioned in Section~\ref{subsec:beam_performance}, the beam features may also have an impact on the waveform features. This was studied by looking at the beam profile data taken using the MWPCs as shown in Fig.~\ref{fig:beamspread}. Fig.~\ref{fig:mwpc}, left, shows the spread of the beam as a function of the beam energy. Here, the beam spread is defined to be the average of the standard deviations computed by fitting a Gaussian distribution to the MWPC data (see Fig.~\ref{fig:beamspread}), averaged over both the X-Y plane, and the 3 MWPCs. This is plotted as a function of the beam energy for an electron trigger in red and a pion trigger in blue. This plot shows that the beam is more focused at the higher energies for both electron and pion data. This is to be expected as it takes more secondary collisions to produce the lower energy beams from the initial 120~GeV primary beam. The beam will spread and the overall flux will also reduce as particles of a lower energy are selected. Indeed, when triggering on pions for energies less than 4~GeV, the flux drops so low that a significant amount of waveforms could not be recorded. This data also shows the spatial spread of the beam entry location. At energies below 4~GeV for electrons, this becomes \textasciitilde20~mm greater (a factor \textasciitilde1.75) than at 4~GeV and above. However, Cherenkov photons from an electron, displaced a distance of \textasciitilde20~mm from the beam entry location would correspond to a timing difference of only \textasciitilde100~ps as detected on the PMT inside the tank. Therefore, we do not expect the spatial spread of the beam at the lower energies to impact the pulse shape. The effect of beam halo increases as the energy is decreased but, as mentioned in Section~\ref{subsec:beam_performance}, this should not be significant due to the requested low flux of the beam.

    Fig.~\ref{fig:mwpc}, right, shows the spread of the primary PMT pulse ($\sigma_p$). The electron beam is shown in red and the pion beam is shown in blue. Unlike the beam monitor data, measured $\sigma_p$ increases for electron data but not pion data. This trend is different from the beam monitor data therefore, we do not see the beam spread having an influence in our measurements.

  \subsection{Pulse shape analysis\label{subsec:pulseshape}}
    \begin{figure}[!b]
      \centering
      \includegraphics[width=\textwidth]{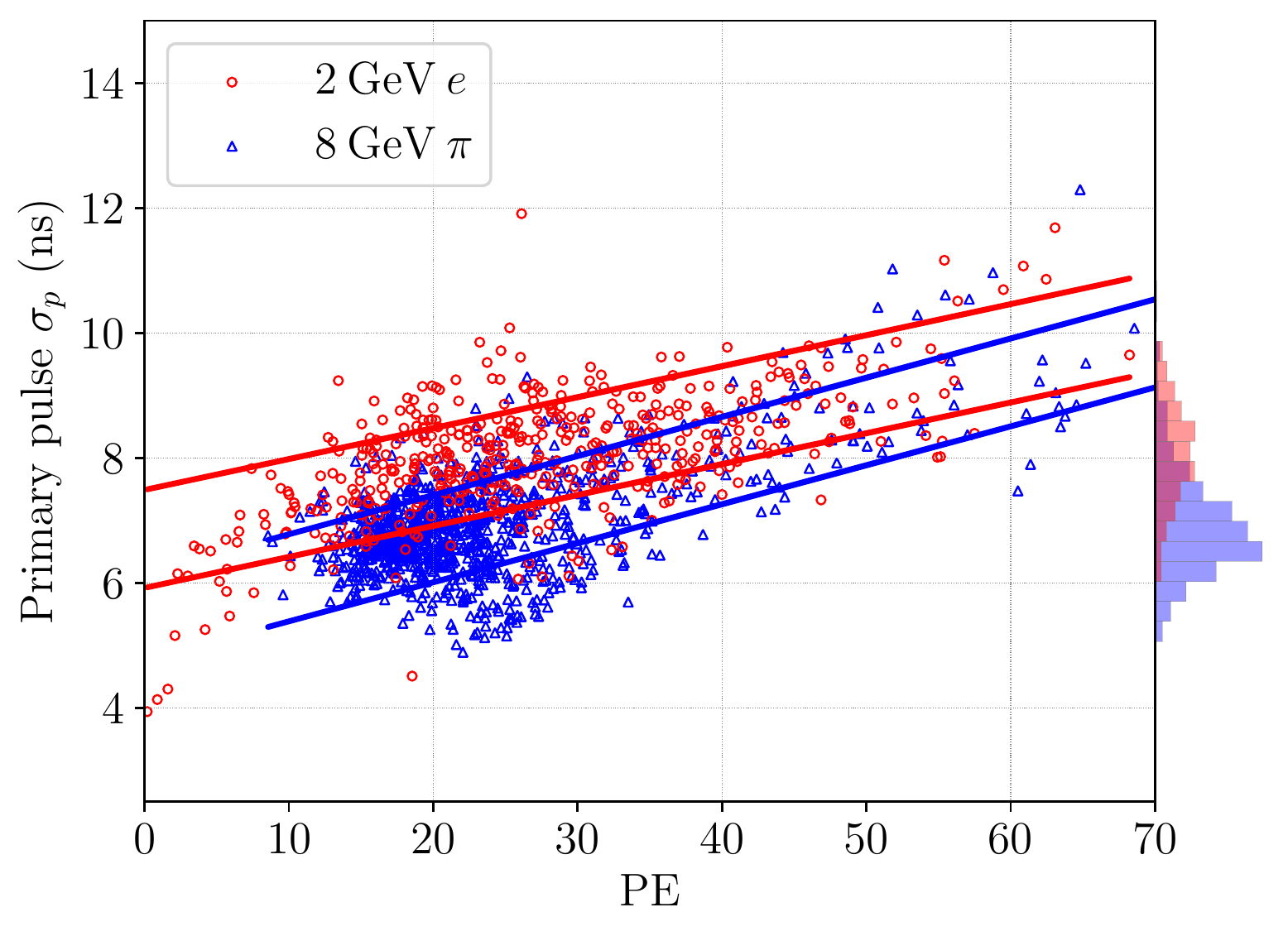}
      \caption{Scatter plot showing the spread of the primary pulse $\sigma_p$ vs the deposited charge. Each red circle corresponds to a waveform from the 2~GeV electron beam and each blue triangle corresponds to a waveform from the 8~GeV pion beam. The red and blue bands show the one sigma containment region for the 2~GeV electron and 8~GeV pion beams, respectively. On the right is shown a histogram of the primary pulse $\sigma_p$ for the respective waveform data.\label{fig:money}}
    \end{figure}
    In the previous sections it is demonstrated that the higher energy beam configurations have problems caused by the growth in the secondary peak and additionally for the electron beam, the saturation of the primary peak. Alternatively, at the lower beam energies the flux becomes small and large statistics cannot be collected. With these points in mind, the 2~GeV electron data has been chosen to compare to the 8~GeV pion data (waveforms shown in Figs.~\ref{fig:waveform-e}~\&~\ref{fig:waveform-pi}). With these datasets, effects such as the secondary pulse, saturation and beam spread are kept to a minimum, while still allowing the comparison of similar charge waveforms. Note, although our {\it in situ} calibration has a problem, the observed charge deposition corresponds to $\sim$20 PE assuming the known gain of this PMT.

    The discriminator that will be used to do the pulse shape analysis here will be the spread of the primary pulse $\sigma_p$. This is shown in Fig.~\ref{fig:money}, where $\sigma_p$ is plotted against the charge, where the red circles represent a waveform from the 2~GeV electron data and the blue triangle a waveform from the 8~GeV pion data. The red band shows the one sigma containment region for the 2~GeV electron data and the blue band shows the one sigma containment region for the 8~GeV pion data. By comparing the two, it can be seen that at the low charges, a statistical discrimination can be made. Here, the EM shower events tend to produce primary pulses which are more spread than for MIP events, as postulated in Section~\ref{subsec:concept}. The significance to which the discrimination can be made is not however to a degree at which the majority waveforms can individually be identified as being from either a MIP or EM shower event. At the higher charges, the power to discriminate is completely lost which can be understood as the result of saturation and secondary pulse effects as previously discussed.

    \begin{figure}[!b]
       \centering
       \includegraphics[width=0.7\textwidth]{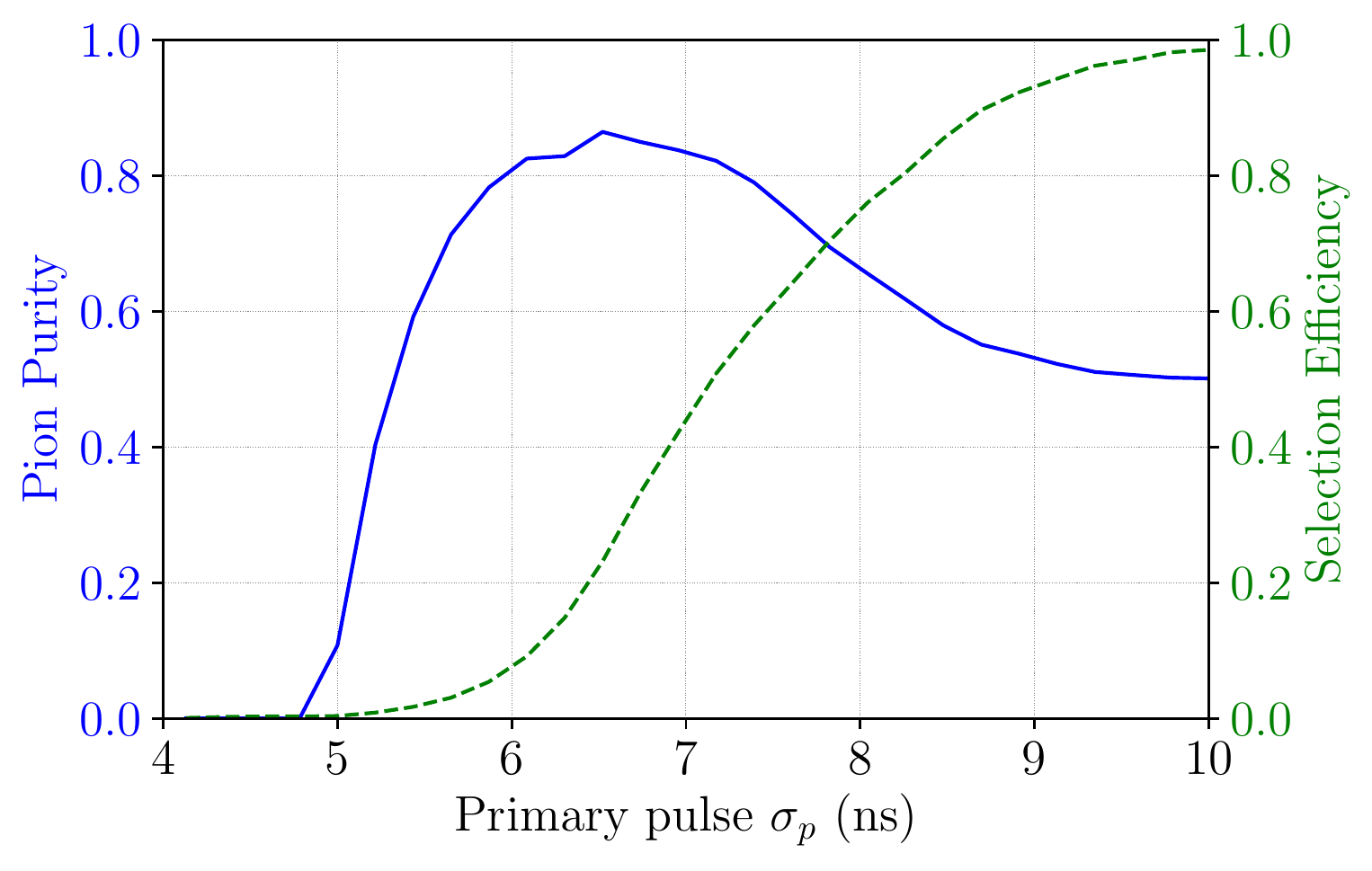} \\
   \caption{\label{fig:PID} Pion purity (solid blue) and selection efficiency (dashed green) as a function of the upper value of the primary pulse width $\sigma_p$ used to select waveforms. The pion purity is normalized assuming an equal mixture of 8~GeV pion and 2~GeV electron data. The figure of merit (purity $\times$ efficiency) can be maximized by choosing a cut at $\sigma_p=7.8$~ns at this beam setting.}
     \end{figure}
    Fig.~\ref{fig:PID} shows an example of a pion selection using the side histogram from Fig.~\ref{fig:money}. The x-axis shows the upper value of the primary pulse width $\sigma_p$ used to select waveforms. The solid blue curve shows the pion purity of this selection, normalized such that the 8~GeV pion and 2~GeV electron samples have equal prior probabilities. These samples deposit a similar number of PE, however it can be seen that they exhibit different primary pulse widths $\sigma_p$, which can be used as a PID variable. The dashed green curve shows the selection efficiency, i.e.\@ the fraction of the total data sample (both pion and electron) which is selected. Based on this figure, we can choose a cut to select pions at $\sigma_p=7.8$~ns to maximize the figure of merit (purity $\times$ efficiency). However, this PID variable is developed using an idealized beam, thus in this beam test we demonstrate the concept of pulse shape particle identification between MIP particles and electromagnetic showers with known trajectories and fixed energies using a single large hemispherical PMT.

\section{Simulation\label{sec:simulation}}
  \subsection{Geant4\label{subsec:geant4}}
    A simulation of this beam test is performed using GEANT4~\cite{Agostinelli:2002hh}, a comprehensive software toolkit designed to simulate physics processes related to particle propagation within matter. Here C++ object-orientated code is utilized to generate a geometrical layout of the experiment and simulate charged particle interactions in the constructed detector. The geometry is setup as follows. The "world" is defined as a $4\times4\times4$~m box which contains all objects. The tank is placed inside this world and its material, dimensions and water content reflect the description given in Section~\ref{subsec:tank}. For the beam test, a 25.4~cm PMT was enclosed in a 35~cm diameter glass shell (see Section~\ref{subsec:dom_spec}). In this simulation, a glass shell with diameter 35~cm is created and placed inside the tank such that the bottom half is immersed in water. The submerged surface of this glass shell is then defined to be the sensitive detector. We consider internal reflections of photons between the water-air surfaces, but not reflections at the surface of the tank wall. This simulation does not include the foam ring, which has the potential to block some photons, or the response of the Tedlar lining which may lead to reflections. We also do not simulate the PMT response, instead, an overall detection efficiency is applied for each photon hitting the sensitive detector. The beam particles are generated upstream and the simulated events are recorded only if the beam particle hits the SC4 scintillator (described in Section~\ref{subsec:triggering}). Note we do not perform a full simulation of the MTest beam line, instead mono-energetic particles are generated in front of the detector assuming that energy losses and spread due to known materials are small.

    Physics processes are then chosen. For this beam test, the GEANT4 libraries for EM physics and muon physics are added, which incorporate processes such as Cherenkov radiation, multiple scattering, Bremsstrahlung and ionization for the electrons and muons, and then pair production, Compton scattering and the photoelectric effect for photons.

    The efficiency and spectral response of the PMT must also be considered. The PMT is specified by Hamamatsu for the wavelength range 300--650~nm~\cite{Aartsen:2016nxy}. However, the optical transmission of the glass falls at wavelengths below 350~nm and at the larger wavelengths the efficiency of the PMT decreases. Therefore, in the simulation, only the Cherenkov photons produced with wavelengths between 350--550~nm are registered as hits on the PMT. The efficiency of the PMT takes into account the quantum efficiency of the PMT, optical absorption in the PMT, glass shell absorption, discriminator threshold effects, and photocathode non-uniformity. Here we will use a flat 10\% efficiency for the photons inside the spectral range specified~\cite{Aartsen:2016nxy}. The simulation is still idealized, in that it does not take into account the SPE spread and the angular dependence of the efficiency of incident photons. Such effects would smear the hit distribution, which we will account for in an ad hoc way using a Gaussian kernel.

  \subsection{Hit distribution\label{subsec:hit_distribution}}
    \begin{figure}[t]
      \centering
      \includegraphics[width=0.7\textwidth]{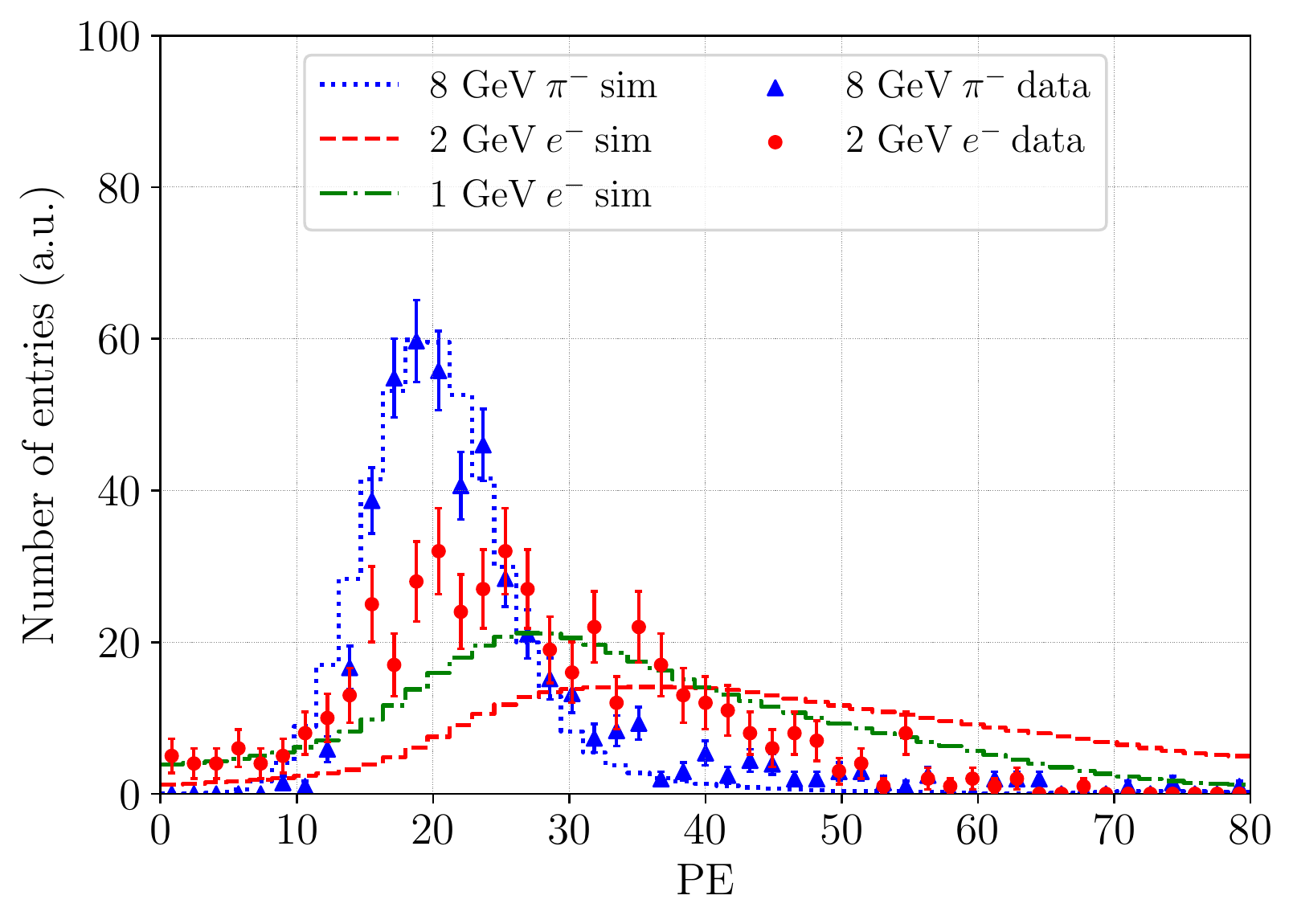}
      \caption{\label{fig:Sim}Hit distribution of the number of photons hitting the PMT for both electron and pion events. Data points taken from this beam test are compared to Geant4 simulations. The presence of saturation can be seen to have an affect for the high charge waveform data, such as for the 2~GeV electron data.}
    \end{figure}
    The hit distribution from data and simulation will be compared to infer how saturation affects the high charge waveform data. Although we estimate efficiencies, we observe a disagreement between data and simulation for the overall hit distribution. Here, we scale the horizontal axis so that the simulated hit distribution peak of the 8~GeV pion data agrees with the simulation. We also apply a smearing on the simulation to again match the pion distribution. We apply the same Gaussian kernel for both pion and electron simulations by assuming the unknown smearing is due to photon propagation physics.

    The results are shown in Fig.~\ref{fig:Sim}. Data is shown for 8~GeV pions as blue triangular points and for 2~GeV electron as circular red points, where both are shown with statistical error bars. The dashed blue line shows the simulation for 8~GeV pions, the dashed red line shows the simulation for 2~GeV electrons and the dashed green line shows the simulation for 1~GeV electrons. The 8~GeV pion simulation is fitted to the data as described above. We apply the same correction to the 1~GeV and 2~GeV electron simulations. Under these conditions, we observe long tails in the electron simulations. This implies that the 2~GeV electron data here is saturated at high charge with the PMT entering the non-linear regime discussed in Section~\ref{subsec:saturation}, causing the voltage read out to plateau. This means that events with high charge are expected to be measured at lower charges as the data suggest. We also observe that the 1~GeV electron simulation has a better agreement with the electron data. This suggests electrons experience energy losses in the beam line through the instrumentation such as the Cherenkov counters, MWPCs, and scintillation counters~\ref{subsec:beam}). We do not expect significant energy loss for MIP particles such as pions.

%\clearpage
\section{Conclusion\label{sec:conclusion}}
Current water Cherenkov telescopes such as IceCube have difficulties identifying particles at low energies (\textasciitilde\@few~GeV) due to the relatively sparse instrumentation and short muon tracks. The low level PMT waveform features are not currently utilized, however the pulse shape shares characteristics of the parent particle. We have demonstrated in this beam test the possibility of performing particle identification using pulse shape analysis of waveforms from a single PMT between MIP-like particles, such as pions, and electromagnetic showering particles, such as electrons. A PMT was floated inside a tank filled with distilled water and using the beam provided by the Fermilab Test Beam Facility, electrons and pions were shot into the tank at different energies. We compare 2~GeV electron and 8~GeV pion waveforms as they deposit similar amounts of charge. The primary pulse spread was used as a discriminator and it is shown that at low charge, there is a statistical discrimination that can be made between 2~GeV electrons and 8~GeV pions. The 2~GeV electron waveforms tend to produce pulses which are more spread when compared to 8~GeV pion waveforms. Since 2-20 GeV is the energy region relevant to neutrino mass ordering measurements, the techniques described in this report could be applied to future neutrino telescopes focusing on low-energy physics, including the DeepCore~\cite{Collaboration:2011ym} and the IceCube-Upgrade~\cite{Ishihara:2019aao, TheIceCube-Gen2:2016cap}.

\acknowledgments

We thank the IceCube collaboration for lending us the DOM and the digitizer used in this beam test. We thank Chris Wendt for his help analyzing the waveforms. We thank Kareem Farrag and Keiichi Mase for their valuable comments to improve this paper. We thank Mandy Rominsky, JJ Schmidt, Todd Nebel, Ewa Skup and Michael Backfish for the great help provided to us at the Fermilab Test Beam Facility. Finally, we thank Sten Hanson, Terry Kiper, Paul Rubinov and Stephen Pordes for providing help with electronics to us at Fermilab. The authors gratefully acknowledge the support from National Science Foundation (NSF-1912764, NSF-1600823), Sun Yat-sen University Young Undergraduates Overseas Immersion Programme, Science and Technology Facilities Council, and the Royal Society International Exchanges.

\bibliographystyle{JHEP}
\bibliography{domtest}

\providecommand{\href}[2]{#2}\begingroup\raggedright\begin{thebibliography}{10}

\bibitem{Aartsen:2016nxy}
{\scshape IceCube} collaboration, M.~G. Aartsen et~al., \emph{{The IceCube
  Neutrino Observatory: Instrumentation and Online Systems}},
  \href{http://dx.doi.org/10.1088/1748-0221/12/03/P03012}{\emph{JINST}
  {\bfseries 12} (2017) P03012},
  [\href{https://arxiv.org/abs/1612.05093}{{\ttfamily 1612.05093}}].

\bibitem{Collaboration:2011ym}
{\scshape IceCube} collaboration, R.~Abbasi et~al., \emph{{The Design and
  Performance of IceCube DeepCore}},
  \href{http://dx.doi.org/10.1016/j.astropartphys.2012.01.004}{\emph{Astropart.
  Phys.} {\bfseries 35} (2012) 615--624},
  [\href{https://arxiv.org/abs/1109.6096}{{\ttfamily 1109.6096}}].

\bibitem{Ishihara:2019aao}
{\scshape IceCube} collaboration, A.~Ishihara, \emph{{The IceCube Upgrade --
  Design and Science Goals}},  in \emph{{HAWC Contributions to the 36th
  International Cosmic Ray Conference (ICRC2019)}}, 2019.
\newblock \href{https://arxiv.org/abs/1908.09441}{{\ttfamily 1908.09441}}.

\bibitem{TheIceCube-Gen2:2016cap}
{\scshape IceCube} collaboration, M.~G. Aartsen et~al., \emph{{PINGU: A Vision
  for Neutrino and Particle Physics at the South Pole}},
  \href{http://dx.doi.org/10.1088/1361-6471/44/5/054006}{\emph{J. Phys.}
  {\bfseries G44} (2017) 054006},
  [\href{https://arxiv.org/abs/1607.02671}{{\ttfamily 1607.02671}}].

\bibitem{Fukuda:1998mi}
{\scshape Super-Kamiokande} collaboration, Y.~Fukuda et~al., \emph{{Evidence
  for oscillation of atmospheric neutrinos}},
  \href{http://dx.doi.org/10.1103/PhysRevLett.81.1562}{\emph{Phys. Rev. Lett.}
  {\bfseries 81} (1998) 1562--1567},
  [\href{https://arxiv.org/abs/hep-ex/9807003}{{\ttfamily hep-ex/9807003}}].

\bibitem{Classen:2019tlb}
{\scshape IceCube} collaboration, L.~Classen, A.~Kappes and T.~Karg, \emph{{A
  multi-PMT Optical Module for the IceCube Upgrade}},  in \emph{{HAWC
  Contributions to the 36th International Cosmic Ray Conference (ICRC2019)}},
  2019.
\newblock \href{https://arxiv.org/abs/1908.10802}{{\ttfamily 1908.10802}}.

\bibitem{Nagai:2019uaz}
{\scshape IceCube} collaboration, R.~Nagai and A.~Ishihara, \emph{{Electronics
  Development for the New Photo-Detectors (PDOM and D-Egg) for
  IceCube-Upgrade}},  in \emph{{HAWC Contributions to the 36th International
  Cosmic Ray Conference (ICRC2019)}}, 2019.
\newblock \href{https://arxiv.org/abs/1908.11564}{{\ttfamily 1908.11564}}.

\bibitem{Aartsen:2019tjl}
{\scshape IceCube} collaboration, M.~G. Aartsen et~al., \emph{{Measurement of
  Atmospheric Tau Neutrino Appearance with IceCube DeepCore}},
  \href{http://dx.doi.org/10.1103/PhysRevD.99.032007}{\emph{Phys. Rev.}
  {\bfseries D99} (2019) 032007},
  [\href{https://arxiv.org/abs/1901.05366}{{\ttfamily 1901.05366}}].

\bibitem{FNAL:FTBF}
``Fermilab test beam facility.'' \url{https://ftbf.fnal.gov/}, 2019.

\bibitem{Rominsky:2018boq}
M.~Rominsky et~al., \emph{{Fermilab Test Beam Facility Annual Report FY17}}, .

\bibitem{Aidala:2017rvg}
{\scshape sPHENIX} collaboration, C.~A. Aidala et~al., \emph{{Design and Beam
  Test Results for the sPHENIX Electromagnetic and Hadronic Calorimeter
  Prototypes}}, \href{http://dx.doi.org/10.1109/TNS.2018.2879047}{\emph{IEEE
  Trans. Nucl. Sci.} {\bfseries 65} (2018) 2901--2919},
  [\href{https://arxiv.org/abs/1704.01461}{{\ttfamily 1704.01461}}].

\bibitem{Hartbrich:2016bbz}
O.~Hartbrich, \emph{Scintillator Calorimeters for a Future Linear Collider
  Experiment}.
\newblock PhD thesis, Bergische U., Wuppertal (main), Hamburg, 2016.
\newblock 10.3204/PUBDB-2016-02800.

\bibitem{TankDepot}
``Tank depot schematics.''
  \url{https://www.tank-depot.com/Drawings%2face%2fop1010-64.pdf}, 2019.

\bibitem{Duvernois:2015pfc}
M.~Duvernois, \emph{{Generation-2 IceCube Digital Optical Module and DAQ}},
  \href{http://dx.doi.org/10.22323/1.236.1148}{\emph{PoS} {\bfseries ICRC2015}
  (2016) 1148}.

\bibitem{Sandstrom:2014mra}
{\scshape IceCube PINGU} collaboration, P.~Sandstrom, \emph{{Digital optical
  module design for PINGU}},
  \href{http://dx.doi.org/10.1063/1.4902801}{\emph{AIP Conf. Proc.} {\bfseries
  1630} (2015) 180--183}.

\bibitem{Abbasi:2008aa}
{\scshape IceCube} collaboration, R.~Abbasi et~al., \emph{{The IceCube Data
  Acquisition System: Signal Capture, Digitization, and Timestamping}},
  \href{http://dx.doi.org/10.1016/j.nima.2009.01.001}{\emph{Nucl. Instrum.
  Meth.} {\bfseries A601} (2009) 294--316},
  [\href{https://arxiv.org/abs/0810.4930}{{\ttfamily 0810.4930}}].

\bibitem{Intel:FPGA}
``Intel cyclone v fpga.''
  \url{https://www.intel.co.uk/content/www/uk/en/products/programmable/fpga/cyclone-v.html},
  2019.

\bibitem{Abbasi:2010vc}
{\scshape IceCube} collaboration, R.~Abbasi et~al., \emph{{Calibration and
  Characterization of the IceCube Photomultiplier Tube}},
  \href{http://dx.doi.org/10.1016/j.nima.2010.03.102}{\emph{Nucl. Instrum.
  Meth.} {\bfseries A618} (2010) 139--152},
  [\href{https://arxiv.org/abs/1002.2442}{{\ttfamily 1002.2442}}].

\bibitem{Kaether:2012bm}
F.~Kaether and C.~Langbrandtner, \emph{{Transit Time and Charge Correlations of
  Single Photoelectron Events in R7081 PMTs}},
  \href{http://dx.doi.org/10.1088/1748-0221/7/09/P09002}{\emph{JINST}
  {\bfseries 7} (2012) P09002},
  [\href{https://arxiv.org/abs/1207.0378}{{\ttfamily 1207.0378}}].

\bibitem{Hamamatsu}
``Photomultiplier tubes basics and applications fourth edition.''
  \url{https://www.hamamatsu.com/resources/pdf/etd/PMT_handbook_v4E.pdf}, 2020.

\bibitem{Agostinelli:2002hh}
{\scshape GEANT4} collaboration, S.~Agostinelli et~al., \emph{{GEANT4: A
  Simulation toolkit}},
  \href{http://dx.doi.org/10.1016/S0168-9002(03)01368-8}{\emph{Nucl. Instrum.
  Meth.} {\bfseries A506} (2003) 250--303}.

\end{thebibliography}\endgroup
%\begin{thebibliography}{99}
% Please avoid comments such as "For a review'', "For some examples",
% "and references therein" or move them in the text. In general,
% please leave only references in the bibliography and move all
% accessory text in footnotes.
% Also, please have only one work for each \bibitem.
%\end{thebibliography}

%\newpage
\appendix
\section{Gain calculation from LED charge distribution\label{sec:gain_calculation}}
  By looking at the charge distribution produced from the PMT using light from a flashing LED, the gain can be estimated, as will be shown in this section. This method is not considered to be accurate however it does yield a quick result, which for our purposes is sufficient to be able to verify the stability of the beam test over time.

  First, we will obtain an estimate of the average number of PE liberated from the photocathode per LED pulse. The distribution we have access to is the charge distribution, so firstly the relationship between the charge and the number of PE can be written as:
  \begin{align}
    Q&=PE\cdot g\cdot C \\
    PE&=\frac{Q}{g\cdot C}=N\cdot Q
  \end{align}
  where $Q$ is the charge, $PE$ is the number of PE, $g$ is the gain, $C$ is the charge on a single electron, i.e.\ $1.6\times10^{-19}$~C and $N$ represents a normalization factor. From this, we see that the number of PE is proportional to the charge.

  The number of PE is distributed as a Poisson distribution, and the charge distribution is related to the PE through a normalization factor $N$. Therefore, by looking at the ratio between the mean $\mu$ and variance $\sigma^2$ of the charge distribution, the average number of PE can be estimated.
  \begin{align}
    \mu=N\cdot m \qquad&\qquad \sigma=N\cdot s \\
    \left(\frac{\mu}{\sigma}\right )^2&=\frac{m^2}{s^2}
  \end{align}
  where $m$ is the mean and $s^2$ is the variance of the PE distribution. Then, from Poisson statistics,
  \begin{align}
    m=s^2 \implies \left(\frac{\mu}{\sigma}\right)^2&=\frac{m^2}{s^2} \\
    \therefore\:\:\:<PE>\:=m&=\left(\frac{\mu}{\sigma}\right )^2
  \end{align}
  Note, throughout these calculations we assume there is no other source contributing to the width of the charge distribution. Since the average number of PE is large, the charge distribution can be approximated as a Gaussian distribution so $\mu$ and $\sigma$ can be obtained from the charge distribution by fitting this, as demonstrated in Fig.~\ref{fig:led_charge}.
  %From here, the gain can be estimated using Ohm's law:
  \begin{align}
    Q=\frac{V\cdot t}{R} \qquad&\qquad V\cdot t = PE\cdot g\cdot C\cdot R\\
    \implies g&=\frac{V\cdot t}{PE\cdot C\cdot R}
  \end{align}
  where $R$ is the impedance, which for the DDC2 is 150~$\Omega$, and $t$ is the time. The product $V\cdot t$ is the mean of the charge distribution.
  \begin{align}
    V\cdot t&=\mu
  \end{align}

  Therefore, the gain can be estimated as being
  \begin{align}
    g=\frac{V\cdot t}{PE\cdot C\cdot R}\implies g=\frac{\mu}{\left(\frac{\mu}{\sigma}\right )^2\cdot C\cdot R}
  \end{align}

% \begin{landscape}
%   \begin{figure}[h!]
%     \vspace*{-50pt}
%     \section{Technical drawings.\label{sec:technical}}
%       \centering
%       \includegraphics[width=1.5\textwidth]{assets/schematics/tank_schematic.pdf}
%       \caption{\label{fig:tank_schematic}TODO(add citation)}
%   \end{figure}
% \end{landscape}

%\begin{figure}[htbp]
%\centering % \begin{center}/\end{center} takes some additional %vertical space
%\includegraphics[width=.4\textwidth,trim=30 110 0 0,clip]{example-image-a}
%\qquad
%\includegraphics[width=.4\textwidth,origin=c,angle=180]{example-image-b}
% "\includegraphics" from the "graphicx" permits to crop (trim+clip)
% and rotate (angle) and image (and much more)
%\caption{\label{fig:i} Always give a caption.}
%\end{figure}

% Please always give a title also for appendices.

\end{document}